\documentclass[a4paper]{aa}
\usepackage{graphicx}
\usepackage{amsmath}
\usepackage{psfig}
\usepackage{natbib}

\bibpunct{(}{)}{;}{a}{}{,}

\def\lsim{\mathrel{\hbox{\rlap{\lower.55ex \hbox {$\sim$}}\kern-.0em
\raise.4ex \hbox{$<$}}}}
\def\gsim{\mathrel{\hbox{\rlap{\lower.55ex \hbox {$\sim$}}\kern-.0em
\raise.4ex \hbox{$>$}}}}

\def\ion#1#2{#1$\;${\small\rm\@Roman{#2}}\relax}
\def\grb{GRB\,050730}

\begin{document}

%

\title{
  Physical conditions in high-redshift GRB-DLA absorbers observed with
  VLT/UVES: Implications for molecular hydrogen searches\thanks{Based
    on Target-Of-Opportunity observations carried out in service mode
    under progs. ID 075.A-0603, P.I. Fiore, and 075.A-0385,
    077.D-0661, 080.D-0526, and 081.A-0856, P.I. Vreeswijk, with the
    Ultraviolet and Visual Echelle Spectrograph (UVES) installed at
    the Nasmyth-B focus of the Very Large Telescope (VLT), Unit~2 --
    Kueyen, operated by the European Southern Observatory (ESO) on
    Cerro Paranal in Chile.}}

\author{
  C.~Ledoux\inst{1}
  \and
  P.~M.~Vreeswijk\inst{2}
  \and
  A.~Smette\inst{1}
  \and
  A.~J.~Fox\inst{1}
  \and
  P.~Petitjean\inst{3}
  \and
  S.~L.~Ellison\inst{4}
  \and
  J.~P.~U.~Fynbo\inst{2}
  \and
  S.~Savaglio\inst{5}
}

\offprints{cledoux@eso.org.}

\institute{
  European Southern Observatory, Alonso de C\'ordova 3107,
  Casilla 19001, Vitacura, Santiago 19, Chile
  \and
  Dark Cosmology Centre, Niels Bohr Institute, University of Copenhagen,
  Juliane Maries Vej 30, 2100 Copenhagen {\O}, Denmark
  \and
  Institut d'Astrophysique de Paris, CNRS and UPMC Paris 6, UMR 7095,
  98bis Boulevard Arago, 75014 Paris, France
  \and
  Department of Physics and Astronomy, University of Victoria,
  Victoria, B.C., V8P 1A1, Canada
  \and
  Max-Planck-Institut f\"ur Extraterrestrische Physik,
  Giessenbachstrasse, PF 1312, 85748 Garching bei M\"unchen, Germany
}

\date{\today}

\authorrunning{Ledoux, Vreeswijk, Smette et al.}
\titlerunning{Physical conditions in GRB-DLAs}


\abstract{}{We aim to understand the nature of the absorbing neutral
  gas in the galaxies hosting high-redshift long-duration gamma-ray
  bursts (GRBs) and to determine their physical conditions.}{A
  detailed analysis of high-quality VLT/UVES spectra of the optical
  afterglow of \grb\ and other {\it Swift}-era bursts is
  presented.}{We report the detection of a significant number of
  previously unidentified allowed transition lines of Fe$^+$,
  involving the fine structure of the ground term ($^6$D$_{7/2}$,
  $^6$D$_{5/2}$, $^6$D$_{3/2}$, $^6$D$_{1/2}$) and that of other
  excited levels ($^4$F$_{9/2}$, $^4$F$_{7/2}$, $^4$F$_{5/2}$,
  $^4$F$_{3/2}$, $^4$D$_{7/2}$, $^4$D$_{5/2}$), from the $z_{\rm
    abs}=3.969$, $\log N($H$^0)=22.10$, damped Lyman-$\alpha$ (DLA)
  system located in the host galaxy of \grb. No molecular hydrogen
  (H$_2$) is detected down to a molecular fraction of $\log f<-8.0$.
  We derive accurate metal abundances for Fe$^+$, S$^+$, N$^0$,
  Ni$^+$, and, for the first time in this system, Si$^+$ and Ar$^0$.
  The absorption lines are best-fit as a single narrow velocity
  component at $z_{\rm abs}=3.96857$. The time-dependent evolution of
  the observed Fe$^+$ energy-level populations is modelled by assuming
  the excitation mechanism is fluorescence following excitation by
  ultraviolet photons emitted by the afterglow of \grb. This UV
  pumping model successfully reproduces the observations, yielding a
  total Fe$^+$ column density of $\log N=15.49\pm 0.03$, a burst/cloud
  distance (defined to the near-side of the cloud) of $d=440\pm 30$~pc,
  and a linear cloud size of $l=520^{+240}_{-190}$~pc. This
  application of our photo-excitation code demonstrates that burst/DLA
  distances can be determined without strong constraints on
  absorption-line variability provided enough energy levels are
  detected. From the cloud size, we infer a particle density of
  $n_{\rm H}\approx 5-15$~cm$^{-3}$.}{We discuss these results in the
  context of no detections of H$_2$ and C\,{\sc i} lines (with $\log
  N($C$^0)/N($S$^+)<-3$) in a sample of seven $z>1.8$ GRB host
  galaxies observed with VLT/UVES. We show that the lack of H$_2$ can
  be explained by the low metallicities, [X/H$]<-1$, low depletion
  factors, and, at most, moderate particle densities of the systems.
  This points to a picture where GRB-DLAs typically exhibiting very
  high H$^0$ column densities are diffuse metal-poor atomic clouds
  with high kinetic temperatures, $T_{\rm kin}\ga 1000$~K, and large
  physical extents, $l\ga 100$~pc. The properties of GRB-DLAs observed
  at high spectral resolution towards bright GRB afterglows differ
  markedly from the high metal and dust contents of GRB-DLAs observed
  at lower resolution. This difference likely results from the effect
  of a bias, against systems of high metallicity and/or close to the
  GRB, due to dust obscuration in the magnitude-limited GRB afterglow
  samples observed with high-resolution spectrographs.}

\keywords{gamma rays: bursts -- galaxies: abundances -- galaxies: ISM --
galaxies: quasars: absorption lines -- cosmology: observations}

\maketitle
%

\section{Introduction}
\label{sec:introduction}

Thanks to prompt response, most notably on 8-10\,m-class ground-based
telescopes, high-resolution optical spectroscopy of rapidly fading
long-duration ($\ga 2$~s) gamma-ray burst (GRB) afterglows provides
detailed information on the kinematics, chemical abundances, and
physical state of the gaseous component of circumburst, host galaxy
interstellar and intergalactic matter intervening these lines of sight
\citep[e.g.,][]{2004A&A...427..785J}. At the GRB host-galaxy redshift,
large amounts of neutral, singly, and highly ionized gas are usually
seen
\citep[e.g.,][]{2005ApJ...624..853F,2007ApJ...666..267P,2008A&A...491..189F},
including the highest neutral hydrogen column densities measured to
date at high redshift from UV absorption-line spectroscopy
\citep[e.g.,][]{2004A&A...419..927V,2005ApJ...634L..25C,2006ApJ...642..979B}.
While such high H$^0$ column densities suggest an origin in the
immediate surroundings of the GRB and/or the inner regions of its
host, the overall H$^0$ column density distribution of the absorbers
is fairly flat \citep{2006A&A...460L..13J}, with values as low as
$\approx 10^{17}$ up to $10^{22.6}$ atoms cm$^{-2}$
\citep{2008ApJ...685..344P,2006ApJ...652.1011W}, illustrating the
diversity of environments probed by the observations. Metallicities
measured in the H$^0$ gas vary widely from less than 1/100$^{\rm th}$
of solar to nearly solar -- with a tendency for low metallicities --
\citep{2006A&A...451L..47F,2007ApJ...666..267P}, consistent with an
origin in dwarf and/or low-mass galaxies and the currently favoured
scenario for long-duration GRBs, the collapse of a massive Wolf-Rayet
star endowed with rotation
\citep{1993ApJ...405..273W,1999ApJ...524..262M,2004IAUS..215..601W}.

Several nearby long-duration GRBs have now been shown to be associated
with supernovae explosions
\citep{1998Natur.395..670G,2003ApJ...591L..17S,2003Natur.423..847H,2004ApJ...609L...5M,2006Natur.442.1011P},
providing the most direct evidence that the progenitors of
long-duration GRBs are indeed massive stars. As most star formation
occurs within molecular clouds, it is generally expected that the
latter are the birthplaces of many GRBs. However, although various
studies \citep{2001ApJ...549L.209G,2002ApJ...565..174R} have argued in
favour of a link between molecular clouds and GRBs, compelling
observational evidence such as the detection of molecular hydrogen
(H$_2$) in GRB optical spectra is lacking
(\citealt{2004A&A...419..927V,2007ApJ...668..667T}). As absorption
lines related to ground-state or vibrationally excited H$_2$ energy
levels should be detectable \citep{2002ApJ...569..780D}, the absence
of detection so far suggests that any H$_2$ molecules in the immediate
vicinity of a GRB are dissociated by the intense X-ray/UV afterglow
flux. However, an H\,{\sc ii} region around the progenitor, or its
star cluster, would also be sufficient to ionize H$^0$ and dissociate
H$_2$ up to distances of 50-100~pc \citep{2008ApJ...682.1114W}.


Unlike intervening QSO absorbers and damped Lyman-$\alpha$ (DLA)
systems ($N($H$^0)\ge 10^{20.3}$ atoms cm$^{-2}$) observed in QSO
spectra, strong absorption lines involving fine-structure and other
metastable levels of ions such as O$^0$, Si$^+$, and Fe$^+$ are
ubiquitous in GRB-DLAs
\citep{2004A&A...419..927V,2005ApJ...634L..25C,2006ApJ...642..979B,2006ApJ...646..358P,2006ApJ...648...95P,2007ApJS..168..231P,2007A&A...467..629D,2007A&A...468...83V,2008A&A...489...37T}.
These levels could be populated by collisions with electrons if the
particle density per unit volume is high enough, by indirect
excitation by IR photons, or by the UV flux from the GRB afterglow. In
\citet{2007A&A...468...83V}, the detection of the time variability of
absorption lines \citep[see also][]{2006ApJ...648L..89D} involving the
fine structure of the ground term and other metastable energy levels
of both Fe$^+$ and Ni$^+$ towards GRB\,060418 was successfully
modelled. This demonstrated that UV pumping followed by de-excitation
cascades is the mechanism at play to form these lines as suggested by
\citet{2006ApJ...648...95P}. The burst/DLA distance inferred in the
above case, $d=1.7$~kpc, represented the first determination of the
distance of the bulk of the absorbing neutral material to a GRB
explosion site. This is consistent with lower limits on the distance,
$d\ga 50$~pc, derived from the observation of strong associated
Mg\,{\sc i} absorption in, e.g., GRB\,051111, which excludes
significant photo-ionization effects \citep{2006ApJ...648...95P}.

Searches for molecular hydrogen in GRB-DLAs, thus equivalently in the
interstellar medium (ISM) of GRB host galaxies, have led to negative
results till now, though in two cases, towards GRB\,050401 and
GRB\,060206, tentative evidence of H$_2$ lines has been reported
\citep{2006ApJ...652.1011W,2006A&A...451L..47F}. The H$_2$
column-density upper limits derived in a handful of cases from the
non-detection of UV absorption lines are somewhat surprising given the
very high H$^0$ column densities observed in some GRB-DLAs. Indeed, in
the Galactic ISM such clouds are found to be nearly fully molecular
\citep[see, e.g.,][]{1977ApJ...216..291S,1979ApJ...231...55J}.
However, several factors could reduce the amount of H$_2$ present in
DLAs: low metallicity, low particle density, strong ambient UV
radiation field, and/or, in the case of GRBs, the UV flux from the
afterglow itself. \citet{2007ApJ...668..667T} argue that the influence
of the latter is negligible and that a combination of low metallicity
and extreme UV radiation field from nearby star-forming regions --
expected to be present in GRB hosts -- must be invoked to explain not
detecting H$_2$ along GRB lines of sight \citep[see
also][]{2008ApJ...682.1114W}. However, given the extremely small
GRB-DLA sample size and the handful of systems considered so far, it
is still unclear whether H$_2$ in GRB host galaxies is really
deficient compared to QSO-DLAs and, if so, why.

In this paper, we present an independent, thorough analysis of
high-quality UVES spectra of \grb\ secured by Fiore et al. (ESO prog.
ID 075.A-0603) at two consecutive epochs after the burst. This line of
sight exhibits one of the highest H$^0$ column density DLAs ever
observed in GRB afterglow spectra
\citep{2005A&A...442L..21S,2005ApJ...634L..25C,2007A&A...467..629D}
with a metallicity at the low end of the distribution for GRB hosts
(see
\citealt{2005ApJ...634L..25C,2007ApJS..168..231P,2007A&A...467..629D};
see also this paper, Sect.~\ref{sec:metallicities}). For the first
time, we identify in the spectrum of this GRB afterglow absorption
lines that originate from numerous Fe$^+$ metastable energy levels.
In addition, we self-consistently model the time-dependent evolution
of the energy-level populations. This allows us to determine the
distance of the bulk of the neutral gas to the GRB explosion site and
the size of the absorbing cloud, and to address the question of
whether the \grb\ afterglow flux is responsible for the observed
stringent upper limit on the H$_2$ molecular fraction. We then discuss
in a more general context the non-detections of H$_2$ molecules taking
advantage of the current sample of seven VLT/UVES spectra of {\it
  Swift}-era, $z>1.8$, GRB host absorbers. We compare them to QSO-DLAs
where H$_2$ has been searched for at high spectral resolution to gain
insights into the physical conditions prevailing in GRB-DLAs.

This paper is organised as follows. In Sect.~\ref{sec:observations},
we recall basic facts on the observations and explain the data
reduction process. In Sect.~\ref{sec:analysis}, we present a detailed
analysis of absorption lines from the GRB host galaxy to constrain the
metallicities, H$_2$ content, and photo-excitation of Fe$^+$ in the
absorbing cloud. We model the time-dependent evolution of the Fe$^+$
metastable energy-level populations in Sect.~\ref{sec:modeling}. In
Sect.~\ref{sec:discussion}, we use H$^0$ photo-ionization and H$_2$
photo-dissociation calculations to assess the influence of the GRB
afterglow radiation and the non-detections of H$_2$ in GRB-DLAs in the
most up-to-date sample of VLT/UVES spectra. This enables us to discuss
the physical nature of GRB-DLAs. Finally, we summarise our findings
and comment on future prospects in Sect.~\ref{sec:conclusions}.

\section{\grb}

\subsection{Observations and data reduction}
\label{sec:observations}

\grb\ was observed with the Ultraviolet and Visual Echelle
Spectrograph \citep[UVES;][]{2000SPIE.4008..534D} mounted at the
Nasmyth-B focus of the ESO VLT, UT~2 -- Kueyen, 8.2\,m telescope on
Cerro Paranal observatory. The data were gathered in service mode
through a classical Target-of-Opportunity request by the programme of
Fiore et al. (ESO prog. ID 075.A-0603). Two $3000$~s exposures were
taken consecutively as soon as possible after evening twilight on July
31, 2005, at 00:32~UT and 1:27~UT (mid-exposure times), corresponding
to, respectively, 4.57 and 5.48~hrs after the burst. The approximate
afterglow I-band magnitude at the mid-exposure time of these two
epochs is $I=17.6$ and $I=18.0$ \citep{2006A&A...460..415P}. Standard
instrument configurations with, respectively, Dichroic \#1 and
Dichroic \#2 were used \citep[both of them using the Blue and Red
spectroscopic arms simultaneously; see also][]{2007A&A...467..629D}.
The first-epoch observations cover the wavelength ranges of 3050 --
3850 and 4800 -- 6800~\AA\ and the second epoch covers the 3850 --
4800 and 6800 -- 10\,000~\AA\ intervals. There are small overlaps
between the two-epoch spectra and, in addition, two $\sim 100$~\AA\
gaps in the Red due to the physical separation between the two Red CCD
detectors. CCD pixels were binned $2\times 2$, and the spectrograph
entrance slit width was fixed to $1\arcsec$ to match the ambient
seeing conditions leading to a resolving power of $\approx
7$~km~s$^{-1}$ (FWHM).

The data were reduced using standard techniques for echelle data
processing with the public version, v2.2.0, of the UVES
pipeline\footnote{http://www.eso.org/projects/dfs/dfs-shared/web/vlt/vlt-instrument-pipelines.html}
based on the ESO MIDAS data reduction package. The main
characteristics of the UVES pipeline are to perform robust inter-order
background subtractions for master flat-fields and science frames, and
to allow for optimal extraction of the object signal performing sky
subtraction and rejecting cosmic rays simultaneously. The pipeline
products were checked step by step. The wavelength scale of the
reduced spectra was then converted to vacuum-heliocentric values. Due
to the variable nature of the source, individual exposures were not
co-added. The typical signal-to-noise ratio per pixel achieved in each
of the two reduced afterglow spectra lies in the range 8 -- 15 over
the 4600 -- 9000~\AA\ interval.

\subsection{Data analysis}
\label{sec:analysis}

The \grb\ line of sight is characterized by the presence of an
extremely strong DLA absorber at $z=3.969$. As no higher redshift
system is observed and Fe\,{\sc ii} lines from metastable levels are
also detected, we assume this is the GRB host-galaxy redshift. The
total neutral hydrogen column density of the \grb-DLA has been
measured in previous analyses
\citep{2005A&A...442L..21S,2005ApJ...634L..25C,2007A&A...467..629D}
and the standard deviation of the three available measurements, based
on different datasets, is very small (0.04 dex). We thus adopt their
average value and combined uncertainty, i.e., $\log N($H$^0)=22.10\pm
0.10$. This is fully consistent with our own measurement based on the
UVES data. Numerous metal absorption lines from neutral, singly, and
highly ionized species associated to the GRB-DLA are detected in the
UVES spectra. While \citet{2007A&A...467..629D} have studied the
multi-component structure of the absorption profiles from a number of
transition lines using the same dataset, we will restrict our analysis
to weak or at most mildly saturated lines. For instance, C\,{\sc
  ii}$\lambda$1334, C\,{\sc ii}$^\star\lambda$1335, O\,{\sc
  i}$\lambda$1302, O\,{\sc i}$^\star\lambda$1304, O\,{\sc
  i}$^{\star\star}\lambda$1306, Si\,{\sc ii}$\lambda$1304, and
Si\,{\sc ii}$^\star\lambda$1309 ground-term fine-structure lines are
detected but they are heavily saturated and, therefore, cannot be used
to derive accurate column density or velocity information.

The overall metal absorption profiles from neutral and singly ionized
species at the GRB host-galaxy redshift are dominated by a single
narrow velocity component at $z_{\rm abs}=3.96857$. Interestingly,
this component alone represents $\ga 90$\% of the total optical depth
of non-saturated lines. This makes the Voigt-profile modelling of
non-saturated lines from the \grb-DLA cloud, as shown below,
exceptionally simple and robust.

\begin{figure}
\centering
\psfig{figure=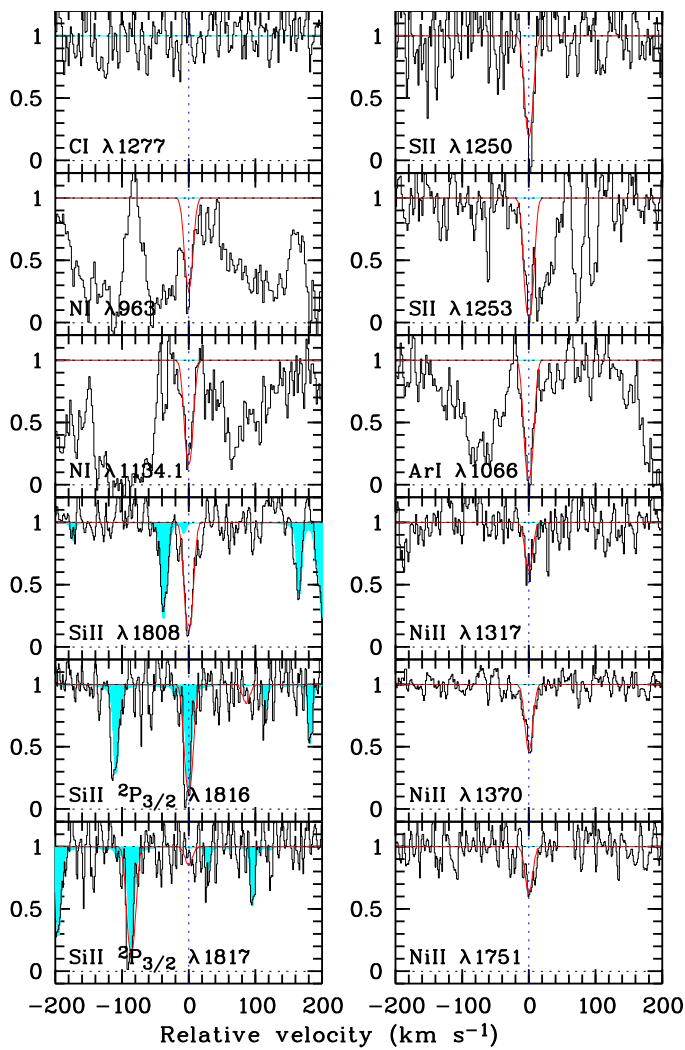,width=8.9cm,clip=,bbllx=61.pt,bblly=237.pt,bburx=403.pt,bbury=768.pt,angle=0.}
\caption{Normalized UVES spectral portions around weaker lines from
  neutral and singly ionized species at the \grb-DLA redshift of
  $z_{\rm abs}=3.96857$ (taken as the origin of the velocity scale).
  Transition lines with rest-frame wavelengths shorter than 1350~\AA\
  were observed at the first epoch and the other lines at the second
  epoch. The best Voigt-profile fit is overplotted. The column density
  from Ni\,{\sc ii}$\lambda$1317, detected in the first-epoch data,
  has been determined independently of the other two Ni\,{\sc ii}
  lines observed at the second epoch. The same turbulent-broadening
  parametre value has been used for all lines and epochs. The shaded
  regions in the panels showing the Si\,{\sc ii} lines indicate the
  expected location and shape of atmospheric features from a synthetic
  telluric absorption-line spectrum described in
  Sect.~\ref{sec:feii}.\label{fig:metals}}
\end{figure}

\begin{table}
\caption{Ionic column densities of neutral and singly ionized species
  at the \grb-DLA redshift of $z_{\rm abs}=3.96857$.\label{tab:ground-state}}
\begin{tabular}{lllcc}
\hline
\hline
Ion          &
Energy level &
Transition   &
${\log N\pm\sigma _{\log N}}^{\rm b}$\\
&& lines used (\AA )&\\
\hline
\multicolumn{4}{l}{1st epoch of observations:}\\
C$^0$  & g.s.$^{\rm a}$                & 1277        &     $<13.05^{\ \rm c}$\\
N$^0$  & g.s.$^{\rm a}$                & 963, 1134.1 &     $14.77\pm 0.03$   \\
S$^+$  & g.s.$^{\rm a}$                & 1250, 1253  &     $15.11\pm 0.04$   \\
Ar$^0$ & g.s.$^{\rm a}$                & 1066        &     $14.36\pm 0.05$   \\
Ni$^+$ & g.s.$^{\rm a}$                & 1317        &     $13.53\pm 0.05$   \\
\hline
\multicolumn{4}{l}{2nd epoch of observations:}\\
Si$^+$ & g.s.$^{\rm a}$ ($^2$P$_{1/2}$) & 1808        &     $15.47\pm 0.03$   \\
Si$^+$ & $^2$P$_{3/2}$                 & 1816, 1817  &     $<15.55^{\ \rm d}$\\
Ni$^+$ & g.s.$^{\rm a}$                & 1370, 1751  &     $13.69\pm 0.02$   \\
\hline
\end{tabular}
\flushleft
$^{\rm a}$ ground state.\\
$^{\rm b}$ Listed uncertainties are the formal errors provided by FITLYMAN.\\
$^{\rm c}$ $3\sigma$ upper limit.\\
$^{\rm d}$ Blend with telluric absorption.
\end{table}

\subsubsection{Metallicities}
\label{sec:metallicities}

We first focus on neutral and singly ionized species excluding Fe$^+$,
which we discuss in Sect.~\ref{sec:feii}. Among previous analyses
\citep{2005ApJ...634L..25C,2007ApJS..168..231P,2007A&A...467..629D},
column density measurements from unsaturated lines were derived by
\citet{2007ApJS..168..231P} for S$^+$, N$^0$, and Ni$^+$. In this
work, we perform simultaneous Voigt-profile fits to the Si\,{\sc ii},
Si\,{\sc ii}$^\star$, S\,{\sc ii}, N\,{\sc i}, Ar\,{\sc i}, and
Ni\,{\sc ii} lines detected in the UVES spectrum and, with the
exception of Si\,{\sc ii}$^\star$ (see below), weak and unblended
enough to allow for an accurate determination of column densities. The
GRB afterglow continuum was normalized locally around the transition
lines of interest using spectral regions with a typical width of
1000~km~s$^{-1}$. In order to fit the lines, the redshift and the
(assumed-to-be) purely turbulent-broadening parametre values were
required to be the same for all lines and the two epochs of
observations. The atomic data compiled by \citet{2003ApJS..149..205M}
were used for all studied transitions, except for the oscillator
strengths of Ni\,{\sc ii}$\lambda\lambda$1317,1370
\citep{2006ApJ...637..548J} and the rest-frame wavelengths of S\,{\sc
  ii}$\lambda\lambda$1250,1253 \citep{1991ApJS...77..119M}. The
spectra and best Voigt-profile fits are shown in Fig.~\ref{fig:metals}
and the results summarised in Table~\ref{tab:ground-state}. The
best-fit turbulent-broadening parametre value is $b=6.5\pm
0.2$~km~s$^{-1}$. The column density measurement uncertainties given
in the table are the formal errors provided by FITLYMAN. They do not
include possible additional uncertainties related to the continuum
placement. Throughout the paper, we adopt the solar system abundances
recommended in \citet{2003ApJ...591.1220L}.

\citet{2005ApJ...634L..25C} estimated the overall metallicity of the
\grb-DLA from the column density of sulphur, $\log N($S$^+)=15.34\pm
0.09$, further refined to be $15.22\pm 0.06$ by
\citet{2007ApJS..168..231P}. Although not considered in the analysis
of \citet{2007A&A...467..629D}, S\,{\sc ii} lines are also detected in
the UVES dataset. In the latter, the 1250~\AA\ feature reaches zero
residual intensity (see Fig.~\ref{fig:metals}), which is not the case
in the MIKE spectra acquired practically simultaneously. While the
UVES spectrum is affected by a narrow spike at the location of the
S\,{\sc ii}$\lambda$1250 line, it is generally of significantly higher
quality in terms of S/N ratio and spectral resolution as can be seen
from the comparison of Fig.~\ref{fig:metals} with figs.~2 and 3 from
\citet{2007ApJS..168..231P} \citep[or equivalently fig.~2
from][]{2005ApJ...634L..25C}. A simultaneous fit of the S\,{\sc
  ii}$\lambda\lambda$1250,1253 lines in the first-epoch UVES spectrum
leads us to derive a column density of $15.11\pm 0.04$, and a
corresponding metallicity of [S/H$]=-2.18\pm 0.11$, somewhat lower
than the previous estimates. The higher estimate by
\citet{2007ApJS..168..231P} is probably caused by the use of the
apparent optical depth method on the blended S\,{\sc ii}$\lambda$1253
line (see Fig.~\ref{fig:metals}).

Our result for sulphur is consistent with the Si$^+$ metallicity we
derive from the second-epoch UVES observations: [Si/H$]=-2.17\pm
0.10$. However, we note that there could be some contribution to the
total Si$^+$ column density from the $^2$P$_{3/2}$ fine-structure
level of the ground term. The expected position of the Si\,{\sc
  ii}$^\star\lambda$1816 line is badly blended with a telluric
absorption feature (see Fig.~\ref{fig:metals}; see also
Sect.~\ref{sec:feii} for details on the telluric-line spectrum).
Therefore, we shall consider in Table~\ref{tab:ground-state} the
result of Voigt-profile fitting of this line strictly as an upper
limit. Given the strength of the blending feature the actual Si$^+$
fine-structure level column density could be at least an order of
magnitude smaller than the value reported in
Table~\ref{tab:ground-state}. The $3\sigma$ column-density upper limit
derived from the non-detection of the much weaker $\lambda$1817 line
is 0.15 dex higher than this value ($<15.7$).

Singly ionized nickel is detected in the UVES spectra at the two
epochs of observations. Ni$^+$ column densities were derived
independently, from the 1317~\AA\ line detected in the first-epoch
spectrum, and from the other two Ni\,{\sc ii} lines
($\lambda\lambda$1370,1751) observed at the second epoch (see
Table~\ref{tab:ground-state}). The resulting values differ by 0.16 dex
($3\sigma$) in the sense of column densities becoming higher with
time. However, given that the uncertainties listed in
Table~\ref{tab:ground-state} do not include errors on the continuum
placement, this result has lower significance. The column density
derived from the Ni\,{\sc ii} line equivalent widths measured by
\citet{2007ApJS..168..231P}, using the sum of their three spectra
taken on average 4.82 hrs (mid-exposure time) after the burst, is
$\log N($Ni$^+)=13.56\pm 0.06$. This is consistent with our
measurements. Finally, we note that, due to the lack of spectral
coverage of the corresponding transition lines, the metastable-level
populations of singly ionized nickel
\citep[see][]{2007A&A...468...83V} cannot be constrained.

Transition lines from neutral nitrogen and neutral argon are detected
in the first-epoch UVES spectrum in a crowded Lyman-$\alpha$ forest
but the lines are narrow and well-defined. We are therefore confident
that the measured column densities (see Table~\ref{tab:ground-state})
are reliable. The N$^0$ column density also agrees with the one
derived by \citet{2007ApJS..168..231P} in the averaged MIKE spectrum.
Finally, the strong C\,{\sc i}$\lambda$1277 transition line is
undetected in the UVES spectrum down to a $3\sigma$ column-density
upper limit of 13.05 (see Fig.~\ref{fig:metals}).

\begin{figure}
\centering
\psfig{figure=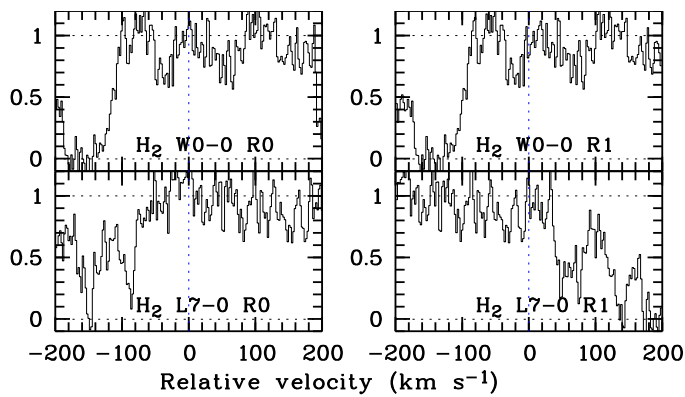,width=8.9cm,clip=,bbllx=61.pt,bblly=567.pt,bburx=403.pt,bbury=768.pt,angle=0.}
\caption{Normalized UVES spectra around several strong transitions to
  the $J=0$ and 1 rotational levels of the vibrational ground-state of
  H$_2$ molecules at the \grb-DLA redshift of $z_{\rm abs}=3.96857$
  (taken as the origin of the velocity scale) in the first-epoch
  observations.\label{fig:H2}}
\end{figure}

\subsubsection{H$_2$ content}
\label{sec:h2}

From the high quality of the UVES data, we can set a stringent upper
limit on the presence of molecular hydrogen in the \grb-DLA. H$_2$ in
its vibrational ground-state is undetected at $3\sigma$ confidence
down to column densities of $\log N($H$_2)=13.30$ and 13.65 for,
respectively, the $J=0$ and 1 rotational levels in the first-epoch
spectrum taken 4.6 hrs (mid-exposure time) after the burst (see
Fig.~\ref{fig:H2}). The expected positions of the Lyman- and
Werner-bands of H$_2$ are not covered by the second-epoch
observations. We used the oscillator strengths from the Meudon
group\footnote{http://amrel.obspm.fr/molat/} based on calculations
described in \citet{1994CanJP..72..856A}. These upper limits translate
to a molecular fraction, $f\equiv
2N($H$_2)/(2N($H$_2)+N($H$^0))<10^{-8.0}$, for the sum of the first
two rotational levels. This is a factor of ten deeper than the
constraint previously derived by \citet{2007ApJ...668..667T} from the
aforementioned MIKE data. This is also the tightest constraint ever
obtained for any extragalactic line of sight at high redshift
including the database of 77 QSO-DLAs from
\citet{2008A&A...481..327N}. We do not find evidence in the UVES
spectra of vibrationally excited H$_2$ lines as would be expected for
dense cold molecular clouds illuminated by incident GRB afterglow
radiation \citep{2002ApJ...569..780D}. This is consistent with the
extremely low molecular fraction derived above from the non-detection
of transition lines to the H$_2$ vibrational ground-state.

\begin{figure}
\centering
\psfig{figure=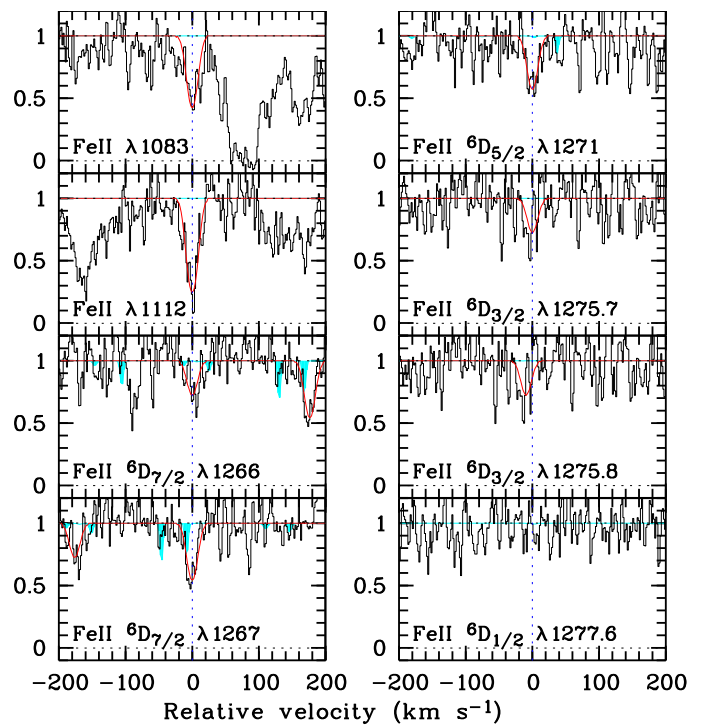,width=8.9cm,clip=,bbllx=61.pt,bblly=402.pt,bburx=403.pt,bbury=768.pt,angle=0.}
\caption{Voigt-profile fitting to transition lines from the
  fine-structure levels of the Fe$^+$ ground-term at the \grb-DLA
  redshift of $z_{\rm abs}=3.96857$ (taken as the origin of the
  velocity scale) in the first-epoch UVES observations. The lower
  level of the transitions, for which the column density is determined
  from the fits, is indicated in each panel, except for the lowest
  Fe$^+$ energy level (ground state) corresponding to $^6$D$_{9/2}$.
  The shaded regions in some of the panels indicate the expected
  location and shape of atmospheric features from a synthetic telluric
  absorption-line spectrum described in
  Sect.~\ref{sec:feii}.\label{fig:epoch1}}
\end{figure}

\begin{figure*}
\centering{\hbox{
\psfig{figure=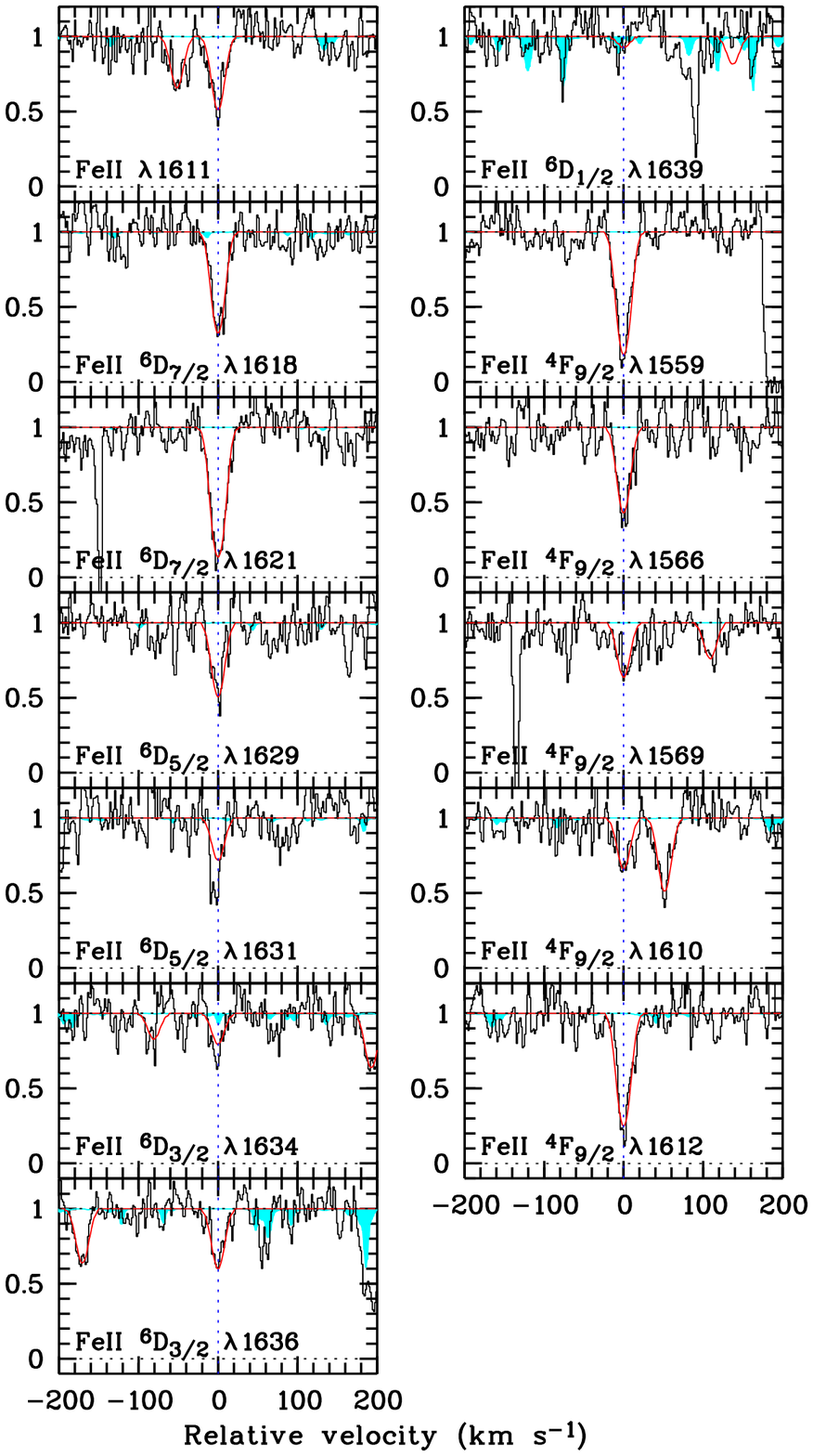,width=8.9cm,clip=,bbllx=61.pt,bblly=155.pt,bburx=403.pt,bbury=768.pt,angle=0.}\hspace{+0.3cm}
\psfig{figure=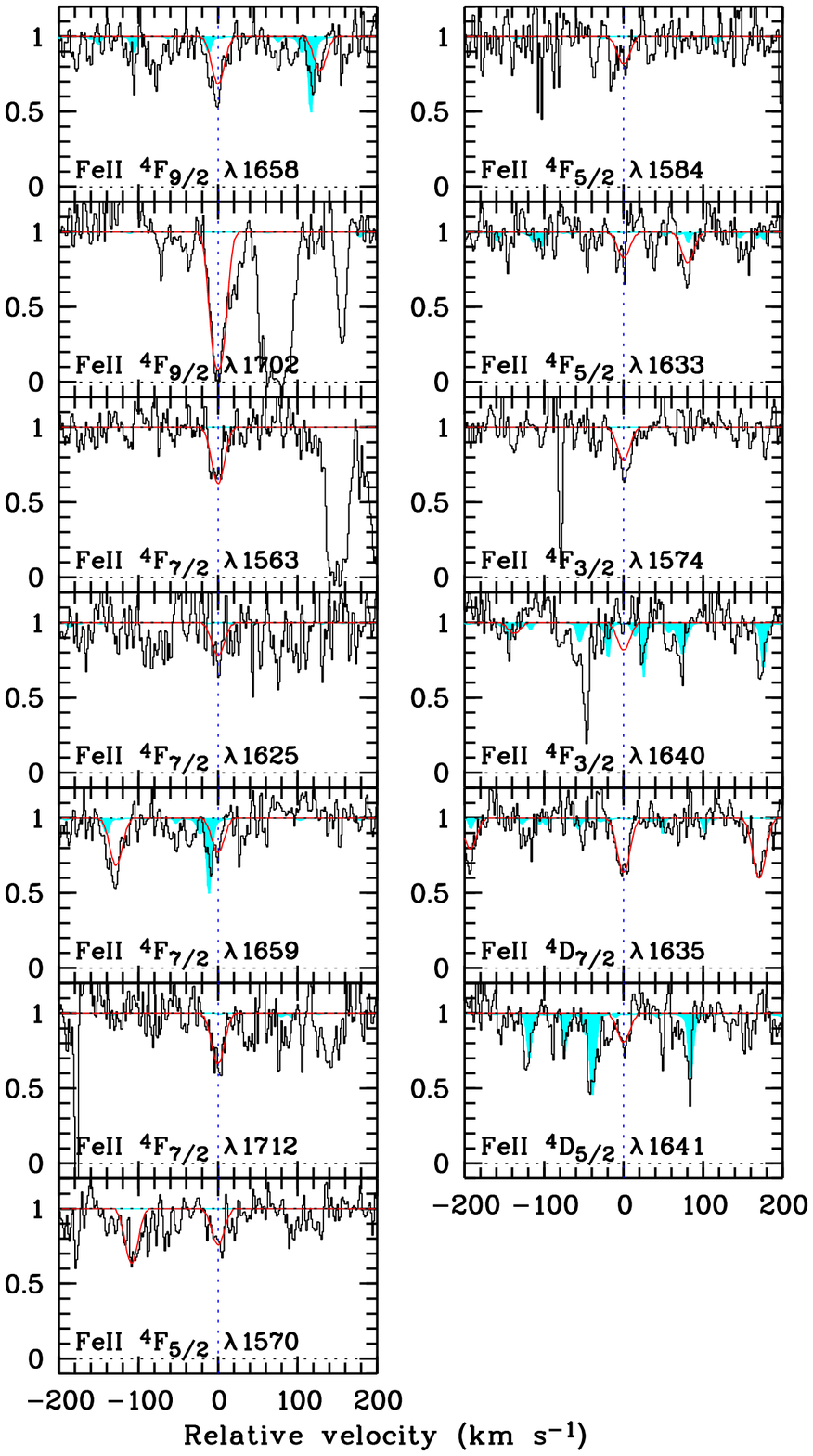,width=8.9cm,clip=,bbllx=61.pt,bblly=155.pt,bburx=403.pt,bbury=768.pt,angle=0.}}}
\caption{Voigt-profile fitting to transition lines from the
  fine-structure levels of the ground term and other metastable levels
  of Fe$^+$ at the \grb-DLA redshift of $z_{\rm abs}=3.96857$ (taken
  as the origin of the velocity scale) in the second-epoch UVES
  observations. The lower level of the transitions, for which the
  column density is determined from the fits, is indicated in each
  panel, except for the lowest Fe$^+$ energy level (ground state)
  corresponding to $^6$D$_{9/2}$. The shaded regions in some of the
  panels, most notably at rest-frame wavelengths of around 1640~\AA,
  indicate the expected location and shape of atmospheric features
  from a synthetic telluric absorption-line spectrum described in
  Sect.~\ref{sec:feii}.\label{fig:epoch2}}
\end{figure*}

\begin{table}
\caption{Column densities of the fine-structure levels of the
  ground term and those of metastable levels of Fe$^+$ at the \grb-DLA
  redshift of $z_{\rm abs}=3.96857$.\label{tab:excited}}
\begin{tabular}{lllc}
\hline
\hline
Ion          &
Energy level &
Transition   &
${\log N\pm\sigma_{\log N}}^{\rm b}$\\
&& lines used (\AA )&\\
\hline
\multicolumn{4}{l}{1st epoch of observations:}\\
Fe$^+$ & g.s.$^{\rm a}$ ($^6$D$_{9/2}$) & 1083, 1112             &     $15.31^{+0.04}_{-0.05}$       \\
Fe$^+$ & $^6$D$_{7/2}$                 & 1266, 1267             &     $14.39^{+0.10}_{-0.12}$       \\
Fe$^+$ & $^6$D$_{5/2}$                 & 1271                   &     $14.36^{+0.09}_{-0.10}$       \\
Fe$^+$ & $^6$D$_{3/2}$                 & 1275.7,  1275.8        &     $14.04^{+0.12}_{-0.16}$       \\
Fe$^+$ & $^6$D$_{1/2}$                 & 1277.64, 1277.68       &    $<13.60^{\ \rm c}$             \\
\hline
\multicolumn{4}{l}{2nd epoch of observations:}\\
Fe$^+$ & g.s.$^{\rm a}$ ($^6$D$_{9/2}$) & 1611                   &     $15.34^{+0.06}_{-0.06}$       \\
Fe$^+$ & $^6$D$_{7/2}$                 & 1618, 1621             &     $14.41^{+0.03}_{-0.03}$       \\
Fe$^+$ & $^6$D$_{5/2}$                 & 1629, 1631             &     $13.96^{+0.06}_{-0.07}$       \\
Fe$^+$ & $^6$D$_{3/2}$                 & 1634, 1636             &     $13.79^{+0.09}_{-0.09}$       \\
Fe$^+$ & $^6$D$_{1/2}$                 & 1639                   &    $<13.65^{\ \rm d}$             \\
Fe$^+$ & $^4$F$_{9/2}$                 & 1559, 1566, 1569       &                                   \\
       &                              & 1610, 1612, 1658       &                                   \\
       &                              & 1702                   &     $14.31^{+0.03}_{-0.04}$       \\
Fe$^+$ & $^4$F$_{7/2}$                 & 1563, 1625, 1659       &                                   \\
       &                              & 1712                   &     $13.64^{+0.10}_{-0.10}$       \\
Fe$^+$ & $^4$F$_{5/2}$                 & 1570, 1584, 1633       &     $13.50^{+0.11}_{-0.14}$       \\
Fe$^+$ & $^4$F$_{3/2}$                 & 1574, 1640             & $\le 13.39^{+0.11\ \rm d}_{-0.12}$\\
Fe$^+$ & $^4$D$_{7/2}$                 & 1635                   &     $13.46^{+0.08}_{-0.08}$       \\
Fe$^+$ & $^4$D$_{5/2}$                 & 1641                   & $\le 13.29^{+0.13\ \rm d}_{-0.15}$\\
\hline
\end{tabular}
\flushleft
$^{\rm a}$ ground state.\\
$^{\rm b}$ Listed uncertainties were derived from the line fitting using the
best-fit continuum normalisation plus or minus 0.5 times the noise RMS
in the adjacent continuum.\\
$^{\rm c}$ $3\sigma$ upper limit.\\
$^{\rm d}$ Possible blend or mis-identification (see text).
\end{table}

\subsubsection{Fe$^+$ ground-state and metastable levels}
\label{sec:feii}

In addition to the study of metal absorption lines commonly observed
in QSO-DLAs (see Sect.~\ref{sec:metallicities}), we have
systematically searched for all absorption features detected at
$3\sigma$ confidence in the UVES spectra and cross-checked their
possible identification using the line lists from
\citet{2003ApJS..149..205M}. To avoid false identifications due to sky
absorption lines, we employed synthetic telluric absorption-line
spectra. The latter were obtained with an IDL routine based on the
Reference Forward Model (RFM)\footnote{http://www.atm.ox.ac.uk/RFM/},
a line-by-line radiative transfer code, using the 2004 edition, v12.0,
of the high-resolution transmission molecular absorption (HITRAN)
database \citep{2005JQSRT..96..139R}. Featured components of RFM
include an atmospheric profile describing mean pressure, temperature,
and molecular concentrations for typically 50 atmospheric layers.
Following \citet{2007EAS....inpressS}, we have calculated synthetic
telluric absorption-line spectra for an amount of precipitable water
vapour of 2.0 mm (resp. 2.4 mm) and a mean airmass of 1.3 (resp. 1.6)
corresponding to the conditions prevailing during the first-epoch
(resp. second-epoch) UVES observations of \grb.

In the line identification process, we uncovered the origin of 18
previously unidentified narrow absorption-line features which are
transition lines from the fine-structure levels of the ground term and
other metastable energy levels of Fe$^+$ at the GRB host-galaxy
redshift of $z_{\rm abs}=3.96857$. This is confirmed by the detection
of typically two or more well-defined lines with consistent optical
depths, and a cross-correlation of their measured rest-frame
wavelengths with the NIST atomic spectra
database\footnote{http://physics.nist.gov/PhysRefData/ASD/index.html}.

In order to fit the lines, we first normalized the spectra around each
line locally in an objective manner using a customized routine which
calculates the median value of the afterglow continuum after rejection
of significant ($\ge 3\sigma$) absorption and emission features. Note
that, in spectral regions heavily affected by telluric absorption,
most notably at rest-frame wavelengths of around 1640 \AA\ (see
below), this procedure tends to underestimate the level of the
continuum. We assumed that the broadening parametre was purely
turbulent. Its value as well as the redshift were assumed to be
identical for all lines and the two epochs of observations. However,
the column densities were allowed to be different for different energy
levels and, for a given energy level, different from one epoch to the
other. The possibility that Fe$^+$ column densities might vary with
time has not been considered in previous analyses of \grb. For
consistency with the time-dependent photo-excitation modelling of the
observed Fe$^+$ level populations presented below, which demands the
inclusion of as many energy levels as possible (see details in
Sect.~\ref{sec:modeling}), we have considered the more complete set of
atomic parametre values from the Cloudy input file based on the
Opacity Project \citep{1996ApJS..103..467V,1999ApJS..120..101V}. In
view of this modelling, care has been exercised to derive column
density measurement errors taking into account the uncertainties in
the continuum placement. These uncertainties indeed dominate the
formal errors provided by FITLYMAN in the case of weak lines. As a
consequence of this, and contrary to Sect.~\ref{sec:metallicities},
errors have been estimated from the line fitting using the best-fit
continuum normalisation (see above) comparing the results with
Voigt-profile fits with the continuum placed at reasonable upper and
lower boundaries. We found these to correspond to adjusting the
continuum level by plus or minus 0.5 times the noise RMS in the
continuum adjacent to the line. The results of Voigt-profile fitting
together with the list of lines used for the fits are shown in
Figs.~\ref{fig:epoch1} and \ref{fig:epoch2}, and summarised in
Table~\ref{tab:excited}. The best-fit turbulent-broadening parametre
value is $b=9.9\pm 1.1$~km~s$^{-1}$. Note that the Fe\,{\sc ii} lines
were fitted separately from the other metal lines resulting in a
somewhat different $b$ value (see Sect.~\ref{sec:metallicities}).
However, since the overall fit is based on weak lines the column
density results are not very sensitive to the exact $b$ value.

Our measurements of column densities for the fine-structure levels of
the Fe$^+$ ground-term ($^6$D) are in most cases consistent with those
derived by \citet{2007A&A...467..629D} and
\citet{2007ApJS..168..231P}. This comparison takes into account the
fact that the oscillator strengths we have adopted differ slightly (by
10-30\%) from those compiled by \citet{2003ApJS..149..205M}, which
were adopted in previous analyses. The only exception to this
consistency is the $^6$D$_{1/2}$ level, whose $\lambda$1639 transition
line is not detected in the second-epoch UVES spectrum (see
Fig.~\ref{fig:epoch2}) while \citet{2007ApJS..168..231P} report a
column density of $\log N=13.65\pm 0.07$. This is 0.65 dex higher than
our $3\sigma$ detection limit, $\log N<13.0$. However, the latter may
be biased low due to our continuum normalisation being affected by the
presence of telluric absorption features, which are conspicuous in
this part of the spectrum as seen from the telluric absorption-line
template (Fig.~\ref{fig:epoch2}). The template also matches the four
absorption-line features at -120, -80, 0, and +80~km~s$^{-1}$ observed
around 8145.5~\AA\ in the MIKE spectrum (see bottom right panel of
fig.~3 in \citealt{2007ApJS..168..231P}), both in terms of wavelength
positions and relative optical depths, pointing to significantly
higher amount of precipitable water vapour present, during the
observations of \grb, in the atmosphere above Las Campanas observatory
than on Paranal. From this, we conclude that the actual Fe$^+$
$^6$D$_{1/2}$ column density must be smaller than 13.65, which we
conservatively adopt as an upper limit in Table~\ref{tab:excited}.

Taken altogether, the column densities measured for the fine-structure
levels of the Fe$^+$ ground-term ($^6$D$_{7/2}$, $^6$D$_{5/2}$, and
$^6$D$_{3/2}$) are consistent with the values becoming smaller with
time. However, the statistical significance of this result is at most
marginal when levels are considered individually. Uncertainties are
large for the first epoch of observations because, at $\lambda _{\rm
  rest}<1300$~\AA, only a few lines are available for a given energy
level, i.e., either a single line or two mutually blended lines (see
Fig.~\ref{fig:epoch1}). It should thus be kept in mind that for the
first epoch some of the fitted lines might be blended or simply
mis-identified. In addition, these lines are weak and noisy. The
$^6$D$_{1/2}$ level is undetected in the first-epoch UVES spectrum
with a $3\sigma$ column-density upper limit of 13.60 (see
Fig.~\ref{fig:epoch1} and Table~\ref{tab:excited}). In contrast, the
second-epoch UVES spectrum displays a wealth of Fe\,{\sc ii} lines
from metastable energy levels. Column densities for the second epoch
of observations are in most cases secured from at least two
well-defined lines, up to seven in the case of the $^4$F$_{9/2}$
level. Exceptions, for which we conservatively quote upper limits or
equal values in Table~\ref{tab:excited}, are the $^4$F$_{3/2}$ level
(for which the feature observed at $\lambda _{\rm rest}\approx
1574$~\AA\ is inconsistent with the non-detection of the $\lambda$1640
line probably because of improper continuum placement) and the
$^4$D$_{5/2}$ level (whose detection through the $\lambda$1641 line is
uncertain due to the presence of adjacent telluric absorption; see
Fig.~\ref{fig:epoch2}).

\begin{figure*}
\centering
\psfig{figure=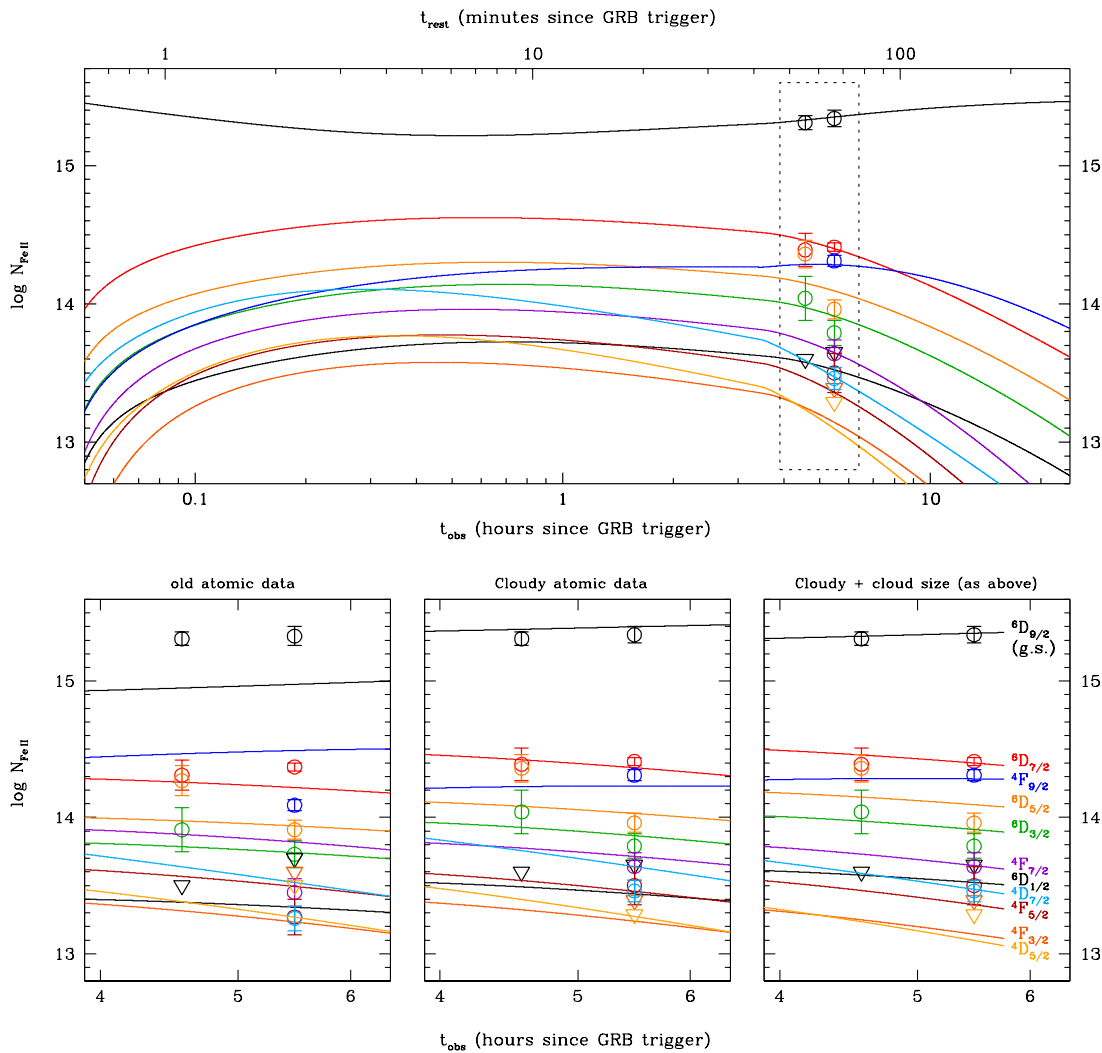,width=15cm,clip=,bbllx=27.pt,bblly=173.pt,bburx=566.pt,bbury=690.pt,angle=0.}
\caption{{\sl Top panel:} Time-dependent evolution of the column
  densities of the ground and several metastable levels of Fe$^+$ in
  our best-fit UV pumping model (indicated with solid lines) using
  Cloudy atomic parametres and allowing for the absorbing cloud to be
  extended (see discussion in text). Measurements (resp. upper limits)
  are indicated with open circles (resp. triangles); the upper limits
  were not used to constrain the fit. {\sl Bottom panels:} Zoom-in
  around the two epochs of observations for three different model
  fits. The bottom right panel shows the same model as the top panel
  but also identifies the Fe$^+$ levels of the observed and modelled
  populations. The other two bottom panels show fits to the data
  assuming a zero cloud size, where the old atomic parametre fit is
  featured in the left panel and the Cloudy parametre fit is located
  in the middle.\label{fig:uv}}
\end{figure*}

\subsection{Fe$^+$ photo-excitation modelling}
\label{sec:modeling}

Our group successfully modelled the time-variable excitation of Fe$^+$
and Ni$^+$ at the host-galaxy redshift of GRB\,060418 as due to
pumping by afterglow ultra-violet (UV) photons
\citep{2007A&A...468...83V}. This led to the first determination of
the distance of the bulk of the host-galaxy absorbing neutral material
to a GRB ($d=1.7\pm 0.2$ kpc). The detection of absorption-line
variability with a steadily rising population of Fe$^+$ in the
$^4$F$_{9/2}$ energy level, and the simultaneous decay of the other
levels, ruled out collisions with electrons as the dominant source of
excitation. There is now mounting evidence that excitation by UV
photons and subsequent de-excitation cascades are the mechanisms at
play to populate the Fe$^+$ metastable energy levels in GRB afterglow
spectra, as variations of Fe\,{\sc ii} absorption lines have been
observed at the host-galaxy redshifts of two additional bursts,
GRB\,020813 and GRB\,080319B, for which suitable multi-epoch data are
available \citep{2006ApJ...648L..89D,2009ApJ...694..332D}. In the case
of \grb, absorption-line variability of transitions involving the
Fe$^+$ $^6$D$_{9/2}$, $^6$D$_{7/2}$, $^6$D$_{5/2}$, and $^6$D$_{3/2}$
levels cannot be firmly established (see Sect.~\ref{sec:feii}) as the
difference between the two epochs of observations is small relative to
the time elapsed since the burst and, therefore, collisions cannot be
excluded as a possible source of excitation. Single-epoch Boltzmann
fits to both the observed ground-state and the metastable level
populations of Fe$^+$, including $^4$F$_{9/2}$ and higher energy
levels, give poor results, with $\chi^2_{\nu}=13.2/(4-2)=6.6$, $T_{\rm
  ex}\sim 400$~K, and $\log N($Fe$^+)=15.3$ for the first epoch, and
$\chi^2_{\nu}=106.8/(8-2)=17.8$, $T_{\rm ex}\sim 2600$~K, and $\log
N($Fe$^+)=14.5$ for the second epoch (while $\log N($Fe$^+)=15.34\pm
0.06$ is observed). The very poor Boltzmann-distribution fits, as well
as the disparate excitation temperatures and ground-state column
densities inferred from the two quasi-simultaneous epoch data, lead us
to reject the hypothesis that collisions are the cause of the
excitation of Fe$^+$. As a consequence, we apply below our UV pumping
model to the case of \grb.

\subsubsection{Model}

In our model, GRB afterglow photons progressively excite Fe$^+$ ions
in a gas cloud located at a distance $d$ from the burst. Although we
take into account both direct excitation by afterglow IR photons and
indirect excitation by UV photons, the latter is, at a given distance,
by far the dominant excitation mechanism
\citep{2006ApJ...648...95P,2007A&A...468...83V}. In addition, as a
result of the large separation of the Fe$^+$ ground-term
fine-structure energy levels, the cosmic microwave background
radiation is always a negligible source of excitation of Fe$^+$ even
at $z=4$ \citep{2002MNRAS.329..135S}. The ion level populations at a
given time are a function of the strength and spectral slope of the
afterglow UV radiation and its decay with time, the pre-burst Fe$^+$
column density (all ions are assumed to be initially in the ground
state), the distance of the absorbing cloud to the GRB explosion site,
and the velocity broadening of the ions. In this respect, our analysis
closely follows that of \citet{2007A&A...468...83V} who modelled the
case of GRB\,060418. Although the narrow time window covered by the
UVES observations of \grb\ does not allow for a definite detection of
absorption-line variability (see above), unlike the UVES observations
of GRB\,060418, the large number of Fe$^+$ metastable energy levels
detected in absorption towards \grb\ constrains the model quite well,
as shown below, so that the burst/DLA distance can be reliably
determined.

The intensity of the UV flux illuminating the cloud is obtained by
converting the observed afterglow light curve to the GRB rest frame.
We used the light-curve description of \citet{2006A&A...460..415P},
namely a broken power law with a break time around 0.1 day with pre-
and post-break power-law indices of $\alpha_1=-0.60\pm 0.07$ and
$\alpha_2=-1.71\pm 0.06$, respectively. These authors also measured a
spectral slope of $\beta=-0.56\pm 0.06$, which we adopted in our
model. \citet{2007arXiv0712.2186K} measured similar values for these
indices. For the flux zero-point of the light curve, we adopted
$I=17.22$ at $t_{\rm obs}$=3.47~hrs after the burst
\citep{2006A&A...460..415P}; this epoch was also assumed to be the
time of the jet break. We note that the $I$-band filter does not
contain the Ly$\alpha$ forest of absorption lines which in turn
affects the $B$, $V$, and $R$-band photometry
\citep[see][]{2006A&A...460..415P}, while the rest-frame central
wavelength of the $I$-band filter is around 1600~\AA, which is in the
wavelength regime (912-2600~\AA) of the UV photons responsible for the
excitation of Fe$^+$. This leads to the following description of the
rest-frame flux at a distance $d$ from \grb:

\begin{equation}
\begin{split}
  F^{\rm rest}_{\nu} = & \frac{3.21\times10^{-27}}{1+z} \cdot\\
  & \left[\frac{t_{\rm obs}}{3.47~{\rm hr}}\right]^{\alpha}
  \left[\frac{\lambda_{\rm obs}}{7977~{\rm\AA}}\right]^{-\beta}
  \left[\frac{3.55\times10^{10}~{\rm pc}}{d}\right]^2
  \label{eq:fnu}
\end{split}
\end{equation}

\noindent in erg s$^{-1}$ cm$^{-2}$ Hz$^{-1}$. In the conversion to
the rest frame, we adopted the Galactic extinction of $E(B-V)=0.046$
from \citet{1998ApJ...500..525S} and assumed that any other extinction
along the line of sight is negligible \citep[see
also][]{2005A&A...442L..21S}. The latter assumption is consistent with
the low metallicity and low depletion factor of [S/Fe] measured at the
host-galaxy absorber redshift, which indicates a low dust content if
any (see Sect.~\ref{sec:metallicities}). For the calculation of the
luminosity distance, $d_l=3.55\times 10^{10}$~pc, we adopted
$H_0=70$~km~s$^{-1}$~Mpc$^{-1}$, $\Omega _{\rm M}=0.3$, and $\Omega
_\Lambda =0.7$. For the calculation of the atom level populations, we
refer the reader to equations 3-5 and the accompanying explanations of
\citet{2007A&A...468...83V}. We checked for the contribution of the
source function by performing two runs including and excluding it; the
change in the resulting chi-square was found to be negligible. As
including the source function requires much more CPU time, we set the
source function to zero. For the velocity broadening of the ions, we
adopted the value that we measured in the UVES spectra,
$b=10$~km~s$^{-1}$ (see Table~\ref{tab:excited}) as it is
well-constrained due to the simplicity of the absorption-line profile
of this GRB-DLA. For many UV transitions, the cloud that we model is
optically thick and, therefore, as in \citet{2007A&A...468...83V}, we
sliced up the cloud in a sufficient number of plane-parallel layers,
so that each layer can be considered optically thin for a given
transition. In contrast to GRB\,060418, we fixed the starting time
$t_0$ to 30~s in the rest frame for GRB 050730 as the results are very
insensitive to any value of $t_0$ between five and 200~s. In other
words, the level populations at roughly an hour (in the rest frame)
after the burst do not depend on the brightness of the very early
afterglow, up to a rest-frame time of about five minutes.

\subsubsection{Atomic data}

We considered two different sets of Fe$^+$ atomic parametres in our
calculations: the first one is the same set as used in
\citet{2007A&A...468...83V}, and the second one is the atom model
employed in the Cloudy photo-ionization code
\citep[see][]{2003ARA&A..41..517F} described in
\citet{1999ApJS..120..101V}. The former set of atomic data, which we
refer to below as the ``old atomic parametres'', includes the 20 lower
energy levels of Fe$^+$ (up to $E=18,886.78$~cm$^{-1}$). The
probabilities for spontaneous decay, or $A$ values, of the forbidden
transitions between all these lower levels were taken from
\citet{1996A&AS..120..361Q}. For the allowed transitions between lower
and higher (starting from $E=38,458.99$~cm$^{-1}$) excited levels, we
adopted the $A$ values compiled by \citet{2003ApJS..149..205M}
whenever available and, if not, we used those provided by
\citet{2003IAUS..210...45K}\footnote{http://kurucz.harvard.edu}. The
Cloudy atomic data set includes all 63 lower energy levels of Fe$^+$.
We here directly cite \citet{1999ApJS..120..101V} for the description
of the corresponding sources: ``Transition probabilities are taken
from theoretical calculations by \citet[][allowed
transitions]{1995A&A...293..967N} and \citet[][forbidden
transitions]{1996A&AS..120..361Q} and supplemented by data from
compilations by \citet{1988atps.book.....F} and
\citet{1995RMxAA..31...23G}. Transition probabilities for all
intercombination lines not covered by these compilations are taken
from the \citet{1995all..book.....K} database. Uncertainties are
generally smaller than 20\% for strong permitted lines but can be
larger than 50\% for weak permitted and intercombination lines. For
the forbidden transitions of interest, uncertainties are expected to
be less than 50\%.'' Using these two different atomic data sets gives
us insight into the uncertainties in the output fit parametres, most
importantly the distance of the photo-excited cloud to the GRB. The
final resulting uncertainties of the model parametres only reflect the
uncertainties in the column density measurements and atomic data used,
and do not include the fact that our model of a plane-parallel series
of absorption slabs is a simplification of the true cloud structure in
the host-galaxy ISM. We also note that the different spontaneous decay
coefficients $A_{\rm ul}$ (or equivalently $f$) of each data set
result in slightly different column densities measured in the spectra.

\subsubsection{Results}

We first ran our photo-excitation code using the old atomic
parametres, which results in a poor fit of the data with a reduced
chi-square of $\chi^2_{\nu}=236.8/(13-3)=23.7$ (see bottom left panel
of Fig.~\ref{fig:uv}). The corresponding best-fit values for the
distance of the cloud to the GRB explosion site and the total Fe$^+$
column density are $d=550\pm 40$~pc and $\log N($Fe$^+)=15.25\pm
0.02$, respectively. When employing the Cloudy atomic parametres, the
fit greatly improves (see bottom middle panel of Fig.~\ref{fig:uv}),
with $\chi^2_{\nu}=24.8/(13-3)=2.5$, and best-fit values of $d=470\pm
30$~pc and $\log N($Fe$^+)=15.52\pm 0.03$. The $1\sigma$ errors
reported here correspond to the parametre value for which the overall
chi-square increases by one with respect to the minimum. This
parametre value is searched for by performing several new fits where
the relevant parametre is fixed, while the other parametres are left
free to vary, until a fit is found where $\Delta\chi^2$ equals one.

The difference between the best fits, obtained using either the old
atomic parametres or the Cloudy atomic parametres, mainly comes from
the very different values of the spontaneous decay coefficients of the
Fe$^+$ $^4$F$_{9/2}$ and $^4$F$_{7/2}$ energy levels. The lower
$A$-values in the old atomic parametre set, by about a factor of
three, cause these levels to be severely overestimated in the model,
as evident from the bottom left panel of Fig.~\ref{fig:uv}. This in
turn results in the populations of the ground state and associated
fine-structure levels of Fe$^+$ being underestimated. Instead, the
Cloudy atomic parametre set results in a very reasonable fit as shown
in the bottom middle panel of Fig.~\ref{fig:uv}. However, we note that
despite the large difference in the atomic data parametre sets used
(both in the number of levels and $A$-values), the derived distances:
$d=550\pm 40$~pc (old set) and $d=470\pm 30$~pc (Cloudy set) are
consistent at the $1.6\sigma$ level.

In contrast to the case of \grb, the observations of GRB\,060418 were
successfully fit using the old atomic parametres
\citep[see][]{2007A&A...468...83V}. Therefore, we verified whether the
time-variability of the excited Fe$^+$ and Ni$^+$ energy-level
populations observed along the line of sight to GRB\,060418 can also
be fit using the Cloudy atomic parametres. We find that the latter
actually provides a much better fit, with a reduced chi-square
($\chi^2_{\nu}=19.3/(36-5)=0.62$) almost half of that of the fit using
the old atomic parametres. The Cloudy best-fit distance of the
absorbing cloud to the GRB ($d=2.0\pm 0.3$~kpc) is fully consistent
within the errors with the previously derived value ($d=1.7\pm
0.2$~kpc), while the best-fit total Fe$^+$ column density is
significantly higher: $\log N($Fe$^+)=15.36\pm 0.11$ (to be compared
to $\log N($Fe$^+)=14.75^{+0.06}_{-0.04}$ from the old fit). This
updated Fe$^+$ column density is consistent with that observed along
the line of sight to GRB\,060418, thereby eliminating the need for
another absorbing cloud, located at much larger distance from the
burst, that is not significantly excited \citep[see the discussion
in][]{2007A&A...468...83V}. This major improvement of the
photo-excitation modelling of the GRB\,060418 data lends strong
additional support for adopting the Cloudy atomic parametres in our
code rather than the old atomic parametres.

Up to this point, we have assumed that all the absorbing neutral
material is located at the same distance along the line of sight to
the GRB, i.e., that the cloud is infinitely small. In an improved
version of our code, we allow the absorbing cloud to have a physical
extent over which the cloud is assumed to have constant density, and
consider its linear size $l$ as an additional fit parametre. Applying
this more realistic model to the data, with each cloud layer (see
above) now located at a different distance from the burst, leads to
the following results: a burst/cloud distance (defined to the
near-side of the cloud) of $d=440\pm 30$~pc, a linear cloud size
(i.e., along the line of sight) of $l=520^{+240}_{-190}$~pc, and a
total Fe$^+$ column density of $\log N($Fe$^+)=15.49\pm 0.03$. It
turns out that the goodness-of-fit improves significantly when
introducing the cloud size, with $\chi^2_{\nu}=12.2/(13-4)=1.4$
(compared to $\chi^2_{\nu}=2.5$ before), and the best-fit parametre
values are well constrained. We will therefore adopt these model fit
results that include a cloud size in the rest of the analysis.
Although in reality the structure of ISM clouds could be much more
complex than our single homogeneous cloud model, the simple
single-component absorption-line profile in the case of GRB\,050730
(see Figs.~\ref{fig:metals}, \ref{fig:epoch1}, and \ref{fig:epoch2})
indicates that our model may actually be a reasonable approximation
for this burst. The fit including a cloud size is shown in both the
top panel and the bottom right panel of Fig.~\ref{fig:uv}. We also
note that the change in light-curve decay index around the break time
of 3.47~hrs is evident in the evolution of the modelled Fe$^+$ level
populations.

\begin{figure*}
\centering{\hbox{
\psfig{figure=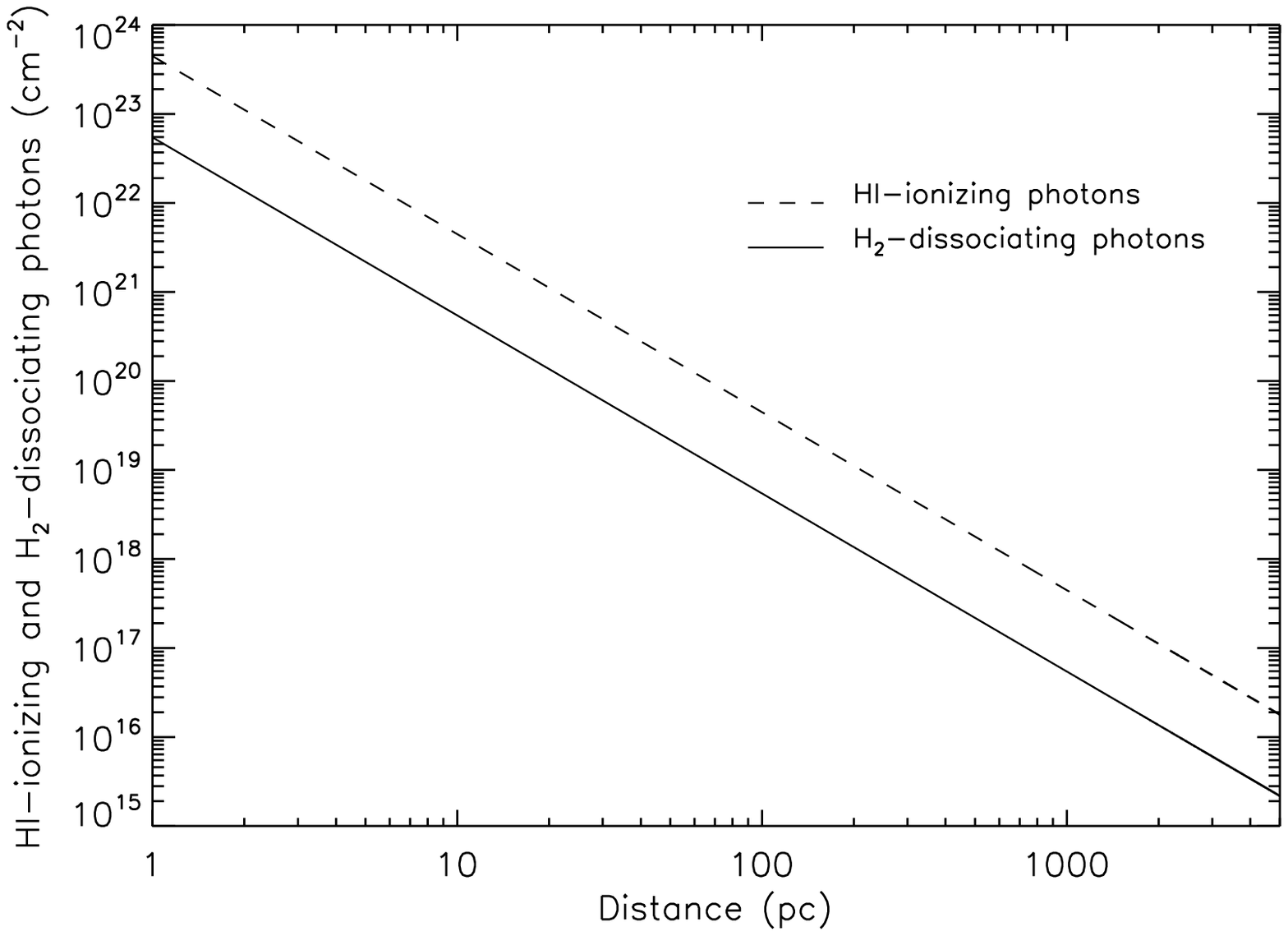,width=8.9cm,clip=,bbllx=76.pt,bblly=369.pt,bburx=538.pt,bbury=706.pt,angle=0.}\hspace{+0.3cm}
\psfig{figure=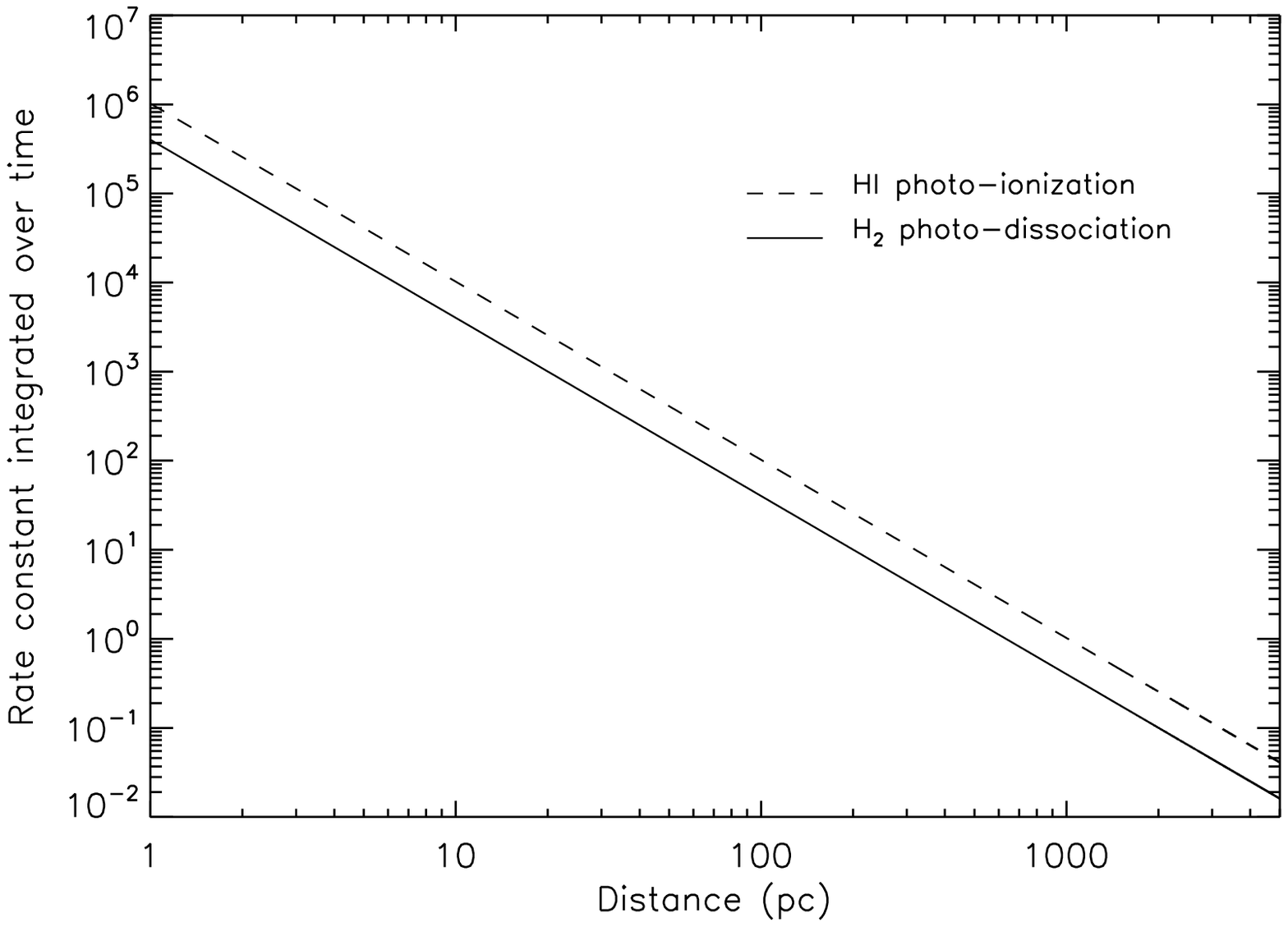,width=8.9cm,clip=,bbllx=76.pt,bblly=369.pt,bburx=538.pt,bbury=706.pt,angle=0.}}}
\caption{{\it Left panel:} Column densities of H$^0$-ionizing and
  H$_2$-dissociating photons released by \grb\ between 30 and 3310~s
  after the burst (rest frame; i.e., up to the mid-exposure time of
  the first-epoch UVES spectrum) as a function of distance to the
  burst. {\it Right panel:} H$^0$ photo-ionization and H$_2$
  photo-dissociation rates integrated over the rest-frame time
  interval 30-3310~s as a function of distance to \grb, neglecting
  self-shielding and assuming no dust (see
  text).\label{fig:dissociation}}
\end{figure*}

\begin{figure*}
\centering{\hbox{
\psfig{figure=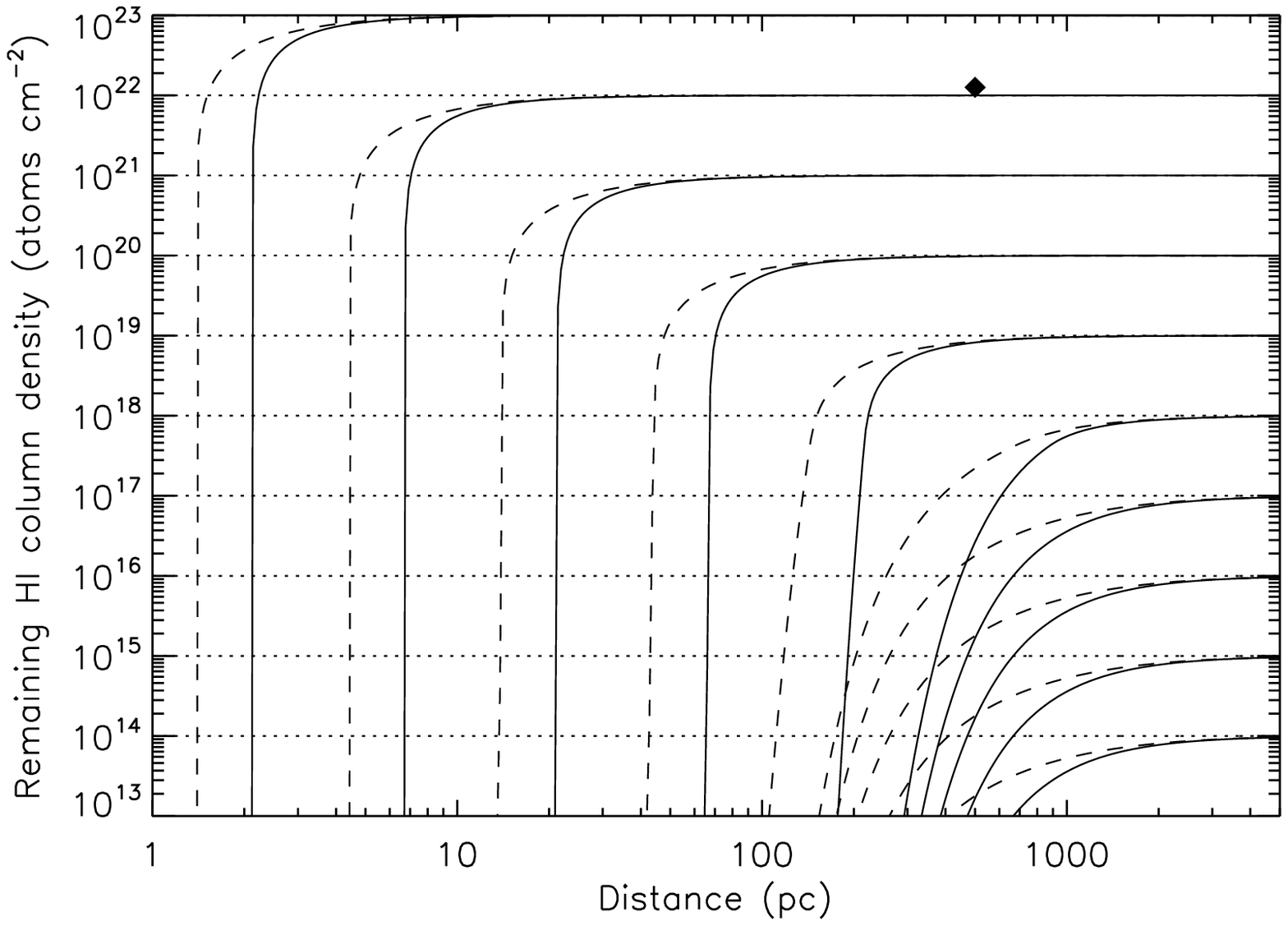,width=8.9cm,clip=,bbllx=76.pt,bblly=369.pt,bburx=538.pt,bbury=706.pt,angle=0.}\hspace{+0.3cm}
\psfig{figure=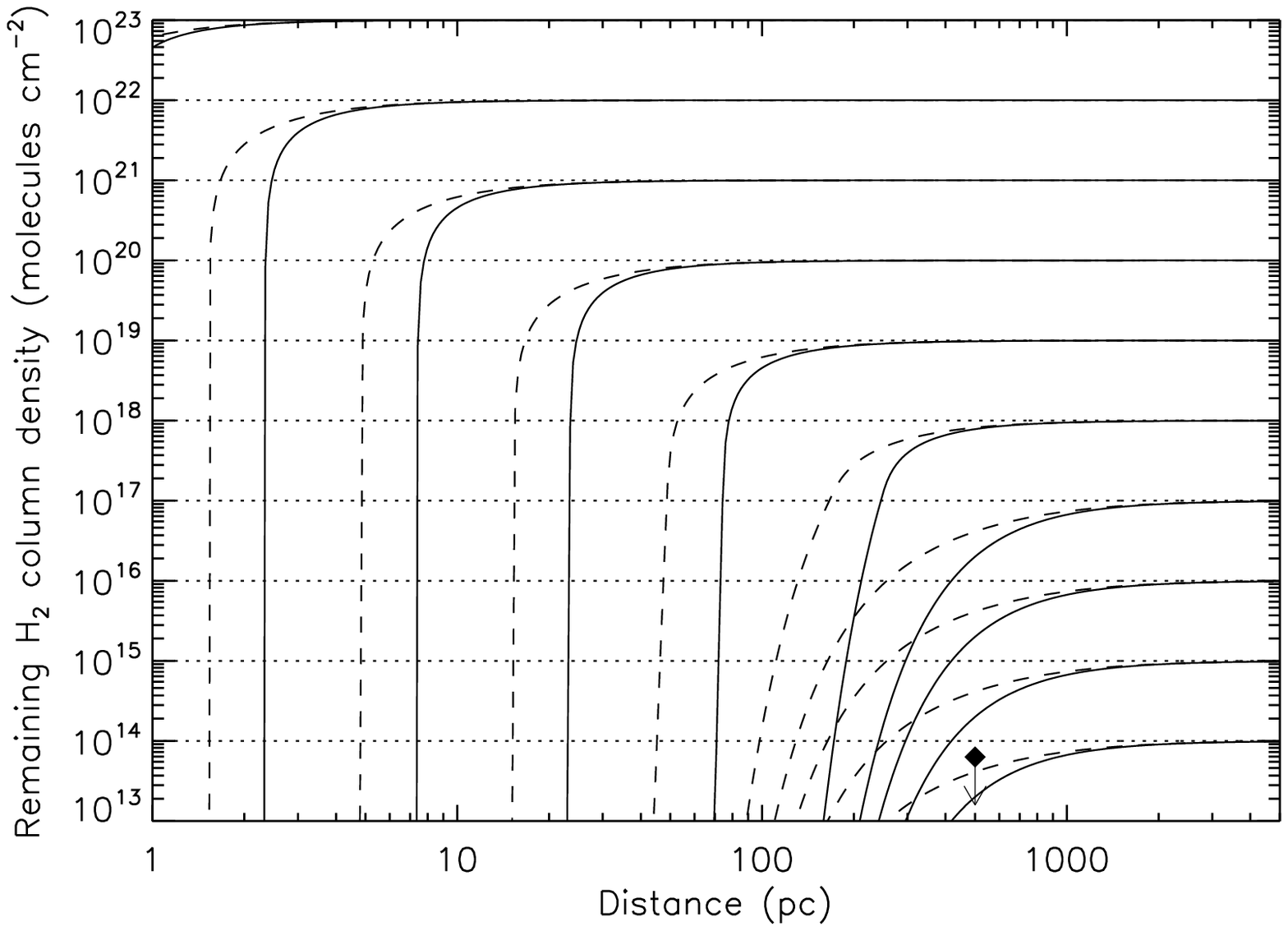,width=8.9cm,clip=,bbllx=76.pt,bblly=369.pt,bburx=538.pt,bbury=706.pt,angle=0.}}}
\caption{Neutral hydrogen (H$^0$; left panel) and molecular hydrogen
  (H$_2$; right panel) column density remaining at the time of the
  first-epoch UVES spectrum as a function of distance to \grb\ for
  dust-free clouds with different initial (pre-burst) column
  densities. The initial column density (10$^{14}$-10$^{23}$) of each
  calculation is indicated with a dotted horizontal line. We performed
  two sets of calculations: one where the cloud size is negligible
  (solid lines), and one where the initial cloud size is assumed to be
  equal to the burst/cloud distance (dashed lines). In each panel, the
  diamond symbol indicates our observational constraint, i.e., $\log
  N($H$^0)=22.10$ on the left and $\log N($H$_2)<13.8$ on the right,
  at a distance of $d\approx 0.5$~kpc from \grb\ as determined from
  the UV pumping model.\label{fig:remainingH2}}
\end{figure*}

\begin{table*}
\caption{Neutral gas, metals, dust, and molecules in the VLT/UVES sample
  of {\it Swift}-era $z_{\rm abs}>1.8$ GRB hosts (as of June
  2008).\label{tab:h2}}
\begin{tabular}{llcclccccc}
\hline
\hline
GRB                      &
$z_{\rm abs}$            &
$\log N($H$^0)$          &
[X/H]                    &
X                        &
[X/Fe$]^{\rm c}$         &
$\log N($Fe$)_{\rm dust}$&
$\log N($H$_2)^{\rm f}$  &
$\log f^{\rm h}$         &
$\log N($C$^0)$          \\
\hline
050730  & 3.969 & $22.10\pm 0.10$ &     $-2.18\pm 0.11$        &            S          & $-0.06\pm 0.06^{\rm d}$ & $<14.77^{\rm d}$ & $<13.8^{\rm g}$ & $<-8.0$ & $<13.05^{\rm g}$\\
050820  & 2.615 & $21.05\pm 0.10$ &     $-0.39\pm 0.10$        &           Zn          & $+0.80\pm 0.03$        &           16.05 & $<14.1^{\rm g}$ & $<-6.7$ & $<12.48^{\rm g}$\\
050922C & 2.200 & $21.55\pm 0.10$ &     $-1.82\pm 0.11$        &            S          & $+0.76\pm 0.05^{\rm e}$ & $<15.12^{\rm e}$ & $<14.6^{\rm g}$ & $<-6.7$ & $<12.45^{\rm g}$\\
060607  & 3.075 & $16.95\pm 0.03$ &        ...$^{\rm a}$        &          ...$^{\rm a}$ &   ...$^{\rm a}$         &    ...$^{\rm a}$ & $<13.5^{\rm g}$ & $<-3.1$ & $<12.03^{\rm g}$\\
071031  & 2.692 & $22.15\pm 0.05$ &     $-1.73\pm 0.05$        &           Zn          & $+0.04\pm 0.02$        &           14.83 & $<14.1^{\rm g}$ & $<-7.8$ & $<12.80^{\rm g}$\\
080310  & 2.427 & $18.70\pm 0.10$ & $\le -1.91\pm 0.13^{\rm b}$ &            O$^{\rm b}$ &   ...$^{\rm a}$         &    ...$^{\rm a}$ & $<14.3^{\rm g}$ & $<-4.2$ & $<12.55^{\rm g}$\\
080413A & 2.435 & $21.85\pm 0.15$ &     $-1.60\pm 0.16$        &           Zn          & $+0.13\pm 0.07$        &           15.13 & $<15.8^{\rm g}$ & $<-5.7$ & $<13.28^{\rm g}$\\
\hline
\end{tabular}
\flushleft
References to the observations: 050730: \citet{2007A&A...467..629D}; 050820:
\citet{2005GCN..3860....1L}; 050922C: \citet{2005GCN..4044....1D};
060607: \citet{2006GCN..5237....1L}; 071031: \citet{2007GCN..7023....1L};
080310: \citet{2008GCN..7391....1V}; 080413A: \citet{2008GCN..7601....1V}.\\
$^{\rm a}$ The metallicity and/or depletion factor cannot be determined due
to ionization effects.\\
$^{\rm b}$ The O\,{\sc i}$\lambda$1302 line used for the fit is partially
blended with strong Al\,{\sc ii}$\lambda$1670 absorption from an intervening
system at $z_{\rm abs}=1.671$.\\
$^{\rm c}$ The total Fe$^+$ column density includes the contributions of
energy levels above the ground state up to $^4$F$_{9/2}$ whenever transition
lines from these levels were detected.\\
$^{\rm d}$ When [X/Fe] is negative, $N($Fe$)_{\rm dust}$ cannot be calculated
directly. In this case, we estimate a $3\sigma$ upper limit on
$N($Fe$)_{\rm dust}$ by considering the upper bound provided by the $3\sigma$
error on [X/Fe].\\
$^{\rm e}$ Significant overabundance of sulphur compared to iron might
be present in this system (see Sect.~\ref{sec:statistics}). The column
density of Fe in dust derived from [S/Fe] (as all measurements based on S or
O) should thus be considered strictly as an upper limit.\\
$^{\rm f}$ Sum of the contributions of the $J=0$ and 1 rotational levels.\\
$^{\rm g}$ $3\sigma$ upper limit.\\
$^{\rm h}$ $f\equiv 2N($H$_2)/(2N($H$_2)+N($H$^0))$.
\end{table*}

\section{Molecular hydrogen in GRB-DLAs}
\label{sec:discussion}

\subsection{GRB-DLA gas distance and H$_2$ photo-dissociation}
\label{sec:distance}

In the previous section, we determined the second precise distance
\citep[but see also][]{2009ApJ...694..332D} from a GRB to the bulk of
the absorbing neutral material experiencing photo-excitation of its
Fe$^+$ component by the time-variable GRB afterglow radiation. We
found that the \grb-DLA is located 0.5~kpc away from the GRB explosion
site. This distance is large enough so that the gas is not strongly
ionized by the incident radiation and, therefore, remains essentially
neutral, in agreement with the observations (see below). This is yet
closer than what we previously determined towards GRB\,060418 at
$z=1.49$, where the burst/DLA distance was found to be 2~kpc
\citep[see Sect.~\ref{sec:modeling} and][]{2007A&A...468...83V}. More
recently, \citet{2009ApJ...694..332D} determined distances of 2-6~kpc
for the different absorption systems in the host galaxy of
GRB~080319B, using the same type of modelling as we presented in
\citet{2007A&A...468...83V} but with different atomic data parametres.
Although it is an open question why the bulk of the absorbing neutral
material is located at such distances from the GRB explosion sites,
these results could be a consequence of the host galaxies of
high-redshift long-duration GRBs being compact, possibly H$^0$-rich,
dwarf galaxies. This would be consistent with the low metallicity of
the \grb-DLA which we measured to be [S/H$]=-2.2$ (see
Sect.~\ref{sec:metallicities}). This would also be in line with the
findings of imaging and spectroscopic studies that the typical
long-duration GRB host at intermediate redshift is a blue,
sub-luminous, low-mass, star-forming galaxy
\citep{2003A&A...400..499L,2004A&A...425..913C,2009ApJ...691..152C,2009ApJ...691..182S}.
In the following, we assess in more detail the influence of the GRB
afterglow radiation on the neutral (H$^0$) and molecular (H$_2$)
contents of a DLA.

Using the observational constraints on the light curve described in
Sect.~\ref{sec:modeling}, we can calculate the 
total number of H$^0$-ionizing photons in units of
s$^{-1}$ cm$^{-2}$ at a given distance from the burst as:

\begin{equation}
n_{\rm phot} = \frac{\int_{\nu_0}^\infty F_\nu \sigma_\nu d\nu /(h\nu )}{\int_{\nu_0}^\infty \sigma_\nu d\nu /\nu }
\label{eq:nphot}
\end{equation}

\noindent where $F_\nu$ is the rest-frame GRB afterglow flux given in
Eq.~\ref{eq:fnu}, $\nu_0$ the frequency of the hydrogen ionisation
threshold (13.6~eV), and $\sigma_\nu$ the H$^0$ photo-ionisation
cross-section. For the latter, we adopt the fits presented in
\citet{1996ApJ...465..487V}. $n_{\rm phot}$ can be integrated over
time from the observed light curve, providing the total column density
of H$^0$-ionizing photons at a distance $d$ released by the GRB. The
left panel of Fig.~\ref{fig:dissociation} (dashed line) shows the
result of this calculation for \grb\ photons emitted between 30 and
3310~s after the burst (rest frame), i.e., up to the time the first
spectrum was taken. In the right panel of Fig.~\ref{fig:dissociation}
(dashed line), we show the H$^0$ photo-ionization rate, or $R_{\rm
  ion}$. It is calculated in the same way as above except that only
the numerator of Eq.~\ref{eq:nphot}, and not its full expression, is
considered: $R_{\rm ion}=\int_{\nu_0}^\infty F_\nu \sigma_\nu d\nu
/(h\nu )$.

In addition, we calculate the total number of H$_2$-dissociating
photons, i.e., the number of photons in the energy range
12.24-13.51~eV (corresponding to the Lyman-Werner absorption bands),
that are released by \grb\ between 30 and 3310~s after the burst
(rest frame). The result is shown as a function of distance to the
burst in the left panel of Fig.~\ref{fig:dissociation} (solid line).
Only 10-15\% of these photons will actually lead to dissociation of
H$_2$ molecules \citep{1996ApJ...468..269D,1999RvMP...71..173H}. The
H$_2$ photo-dissociation rate is given by \citep[see,
e.g.,][]{1997NewA....2..181A}:

\begin{equation}
R_{\rm diss} \approx 1.1\times 10^8\ J_{\rm LW}\ S_{\rm shield}\ {\rm s}^{-1}
\label{eq:rdiss}
\end{equation}

\noindent where $J_{\rm LW}$ (in units of erg s$^{-1}$ cm$^{-2}$
Hz$^{-1}$) is the rest-frame GRB afterglow flux at mean energy
$h\bar\nu =12.87$ eV, and $S_{\rm shield}$ a correction factor for
H$_2$ self-shielding and destruction of UV photons by dust grains
(\citealt{1996ApJ...468..269D}; see also
\citealt{2005MNRAS.356.1529H}). In the right panel of
Fig.~\ref{fig:dissociation} (solid line), we plot $R_{\rm diss}$, or
equivalently $k_{27}$ \citep{1997NewA....2..181A}, integrated over the
rest-frame time interval 30-3310~s as a function of distance to the
burst. We neglect self-shielding and assume no dust as the most
conservative approach to reveal the maximum influence of the
propagating GRB afterglow radiation. It can be seen from this figure
that $R_{\rm diss}=1$ at $d\approx 500$~pc and, therefore, that H$_2$
photo-dissociation can only be effective at $d<500$~pc.

From the definition of the rate constant, i.e., $dN = R N dt$, we now
calculate the H$^0$ and H$_2$ column densities remaining at the time
the first \grb\ UVES spectrum was taken, i.e., 3310~s after the onset
of \grb\ (rest frame), for clouds located at given distances to the
burst and having different initial (pre-burst) column densities (see
Fig.~\ref{fig:remainingH2}).
Each calculation starts out by placing a cloud with an assumed initial
column density at large distance, where H$^0$ ionization and H$_2$
dissociation are negligible. As the distance decreases, the cloud is
increasingly affected by the afterglow photons, until the distance is
reached where the entire column density is ionized or dissociated.
This is shown in Fig.~\ref{fig:remainingH2} for both H$^0$ ionization
(left panel) and H$_2$ dissociation (right panel). The two different
line types depicted in Fig.~\ref{fig:remainingH2} correspond to two
different assumptions for the cloud size: one assumes that the cloud
size is negligible (solid lines), while the other assumes that the
initial cloud size is equal to the burst/cloud distance. The
calculations take into account the fact that the column density of
photo-ionized (or photo-dissociated) particles never exceeds the
column density of photons available at a given distance. We here again
conservatively neglect self-shielding and, in addition, assume no dust
is present, which is relevant to the case of \grb.

One can see in Fig.~\ref{fig:remainingH2} that, at typical distances
of 0.5~kpc and larger, incident GRB afterglow radiation cannot
significantly ionize H$^0$ (left panel) nor dissociate H$_2$ (right
panel) within a DLA. For a DLA cloud, i.e., with $\log
N($H$^0$$)>20.3$, significant ionization effects can only occur at
distances smaller than 50~pc. Moreover, at distances larger than
0.5~kpc the decrease in H$_2$ column density is smaller than 0.4 dex.
Including the process of destruction of UV photons by dust grains
would decrease the influence of the GRB even more. However, this will
be important only for high $N($H$^0)$ values coupled to high
metallicities and large dust contents, while for \grb\ a low dust
content was inferred \citep[see][]{2005A&A...442L..21S}. Therefore,
the current lack of H$_2$ detections in GRB-DLAs located at least
0.5~kpc away from the GRB explosion sites must be caused by one or
several factors other than the propagating afterglow radiation
\citep[see also][]{2007ApJ...668..667T}. This is what we investigate
below by considering the most up-to-date sample of high-resolution
VLT/UVES spectra of high-redshift GRB afterglows.

\begin{figure*}
\centering{\hbox{
\psfig{figure=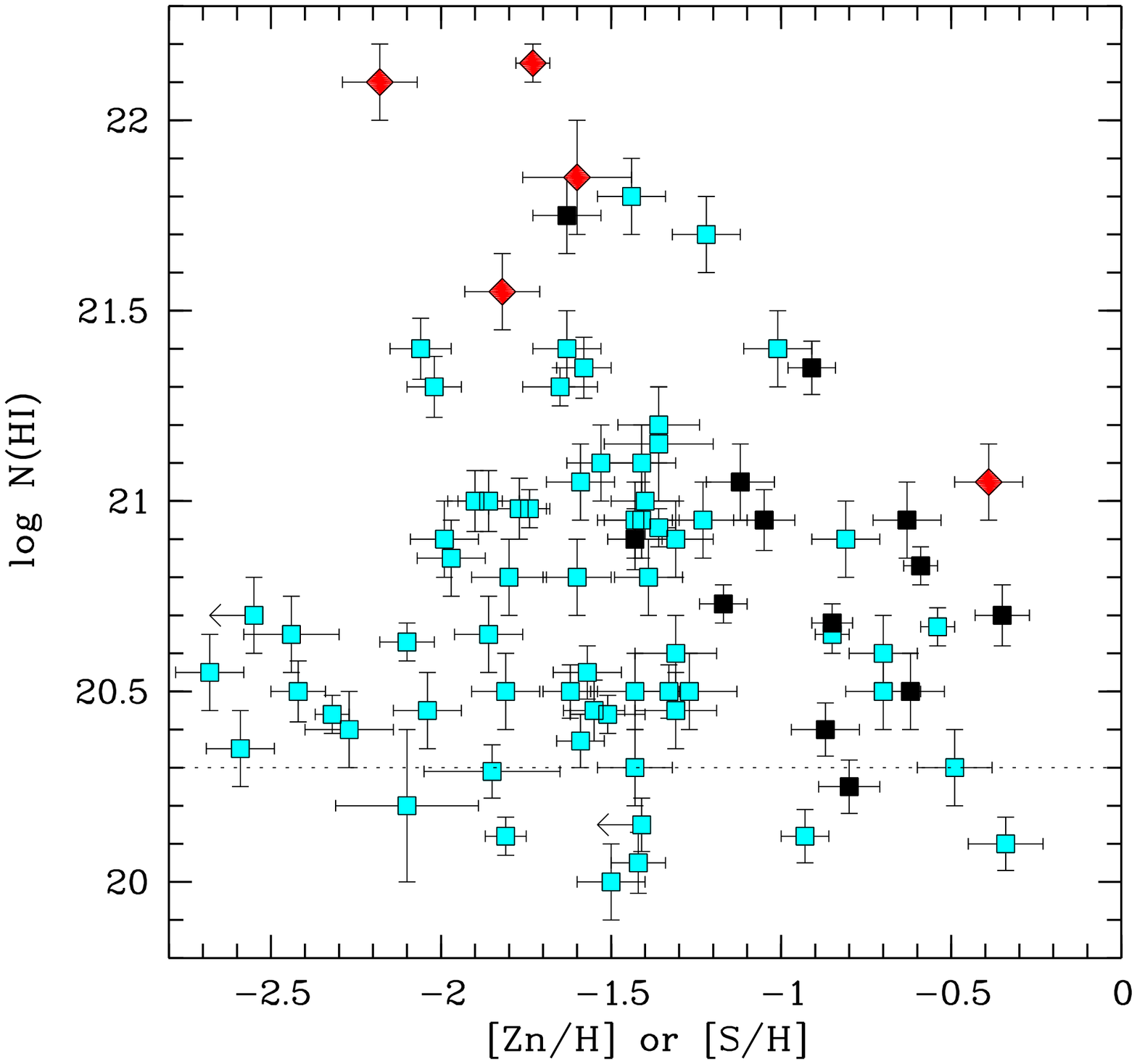,width=8.9cm,clip=,bbllx=52.pt,bblly=65.pt,bburx=563.pt,bbury=611.pt,angle=-90.}\hspace{+0.3cm}
\psfig{figure=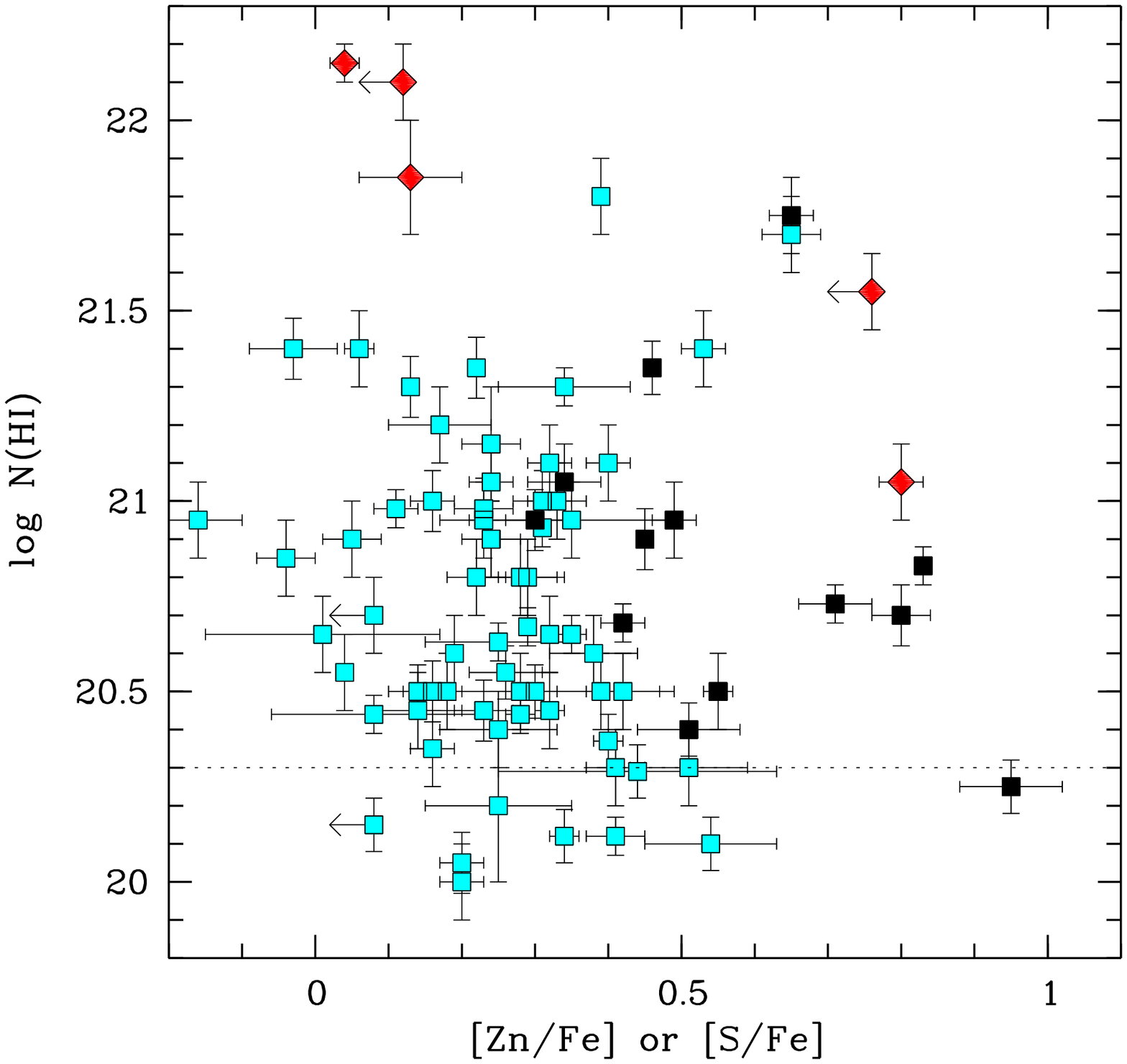,width=8.9cm,clip=,bbllx=52.pt,bblly=65.pt,bburx=563.pt,bbury=611.pt,angle=-90.}}}
\caption{{\it Left panel:} Logarithm of the neutral hydrogen column
  density versus metallicity, [X/H$]\equiv\log N($X$)/N($H$)-\log
  ($X$/$H$)_\odot$, with X$=$Zn or S (see text) in the UVES GRB-DLA
  absorber sample (red diamonds). Similar measurements in the sample
  of QSO-DLAs observed with UVES \citep{2008A&A...481..327N} are shown
  with squares. Black squares are for H$_2$-bearing QSO-DLAs and blue
  ones for H$_2$ non-detections. There is no H$_2$ detection in the
  GRB-DLA sample. {\it Right panel:} Same as before but as a function
  of depletion factor, [X/Fe]. Due to their very low H$^0$ column
  densities, GRB~060607 and GRB~080310 are not featured in these
  plots.\label{fig:HImetdep}}
\end{figure*}

\subsection{Statistics of H$_2$ in GRB-DLAs}
\label{sec:statistics}

In Table~\ref{tab:h2}, we list the seven {\it Swift}-era GRB
afterglows with redshifts higher than 1.8 for which UVES observations
were secured as of June 2008.
For each line of sight, we measured in this work total H$^0$ column
densities from Voigt-profile fitting to the Ly$\alpha$ and/or
Ly$\beta$ absorption lines at the GRB host-galaxy redshifts (denoted
$z_{\rm abs}$ in the table). In the same way as described in
Sect.~\ref{sec:metallicities}, we have also measured or re-measured in
an homogeneous manner, from Voigt-profile fitting to the associated
metal absorption lines, metallicities, [X/H$]\equiv\log
N($X$)/N($H$)-\log ($X$/$H$)_\odot$, and depletion factors, [X/Fe],
where X is a non-refractory reference element. X is taken to be Zn
when Zn\,{\sc ii} lines are detected, or S otherwise. For GRB hosts
that do not meet the requirement for DLA absorbers ($\log
N($H$^0)<20.3$), the abundance of neutral oxygen may be used instead.
This atom is linked to neutral hydrogen by a strong charge-exchange
reaction due to the similarity of ionization potential
\citep{1971ApJ...166...59F}, and the [O$^0$/H$^0$] ratio can lead to
an estimate of the total oxygen abundance \citep{1995MNRAS.276..268V}.
The ionization correction factor is smaller than 10\% for GRB\,080310
having a logarithmic H$^0$ column density of 18.8. However, for
GRB\,060607 with $\log N($H$^0)=16.95$ the ionization correction
highly depends on the hydrogen particle density and, therefore, no
robust metallicity estimate can be derived. To measure depletion
factors and get a handle on the dust content of the systems, iron was
selected as a proxy for refractory elements. Total Fe$^+$ column
densities were calculated by summing up the contributions of both the
ground state and the metastable energy levels of Fe$^+$ when detected.
The contribution of the latter levels never exceeds 0.2 dex. In
addition, we note that the corresponding absorption lines are not
detected at all in the DLAs towards GRB\,050820 and GRB\,080413A.

H$_2$ is not detected in any of the systems of the sample. In
Table~\ref{tab:h2}, we give $3\sigma$ upper limits on the molecular
hydrogen column densities and corresponding H$_2$ molecular fractions,
$f\equiv 2N($H$_2)/(2N($H$_2)+N($H$^0))$. We also give $3\sigma$ upper
limits on $N($C$^0)$.

Apart from the DLA towards GRB\,050820 where the metallicity is as
high as [Zn/H$]=-0.4$, the GRB host galaxies in the UVES sample have
low metallicities, $-2.2<[$X/H$]<-1.3$. In addition, the depletion
factors are usually quite small, as expected for metal-poor
environments having a low dust content. The only exception to this is
the GRB\,050922C-DLA for which sulphur is used as a proxy for
non-refractory elements. In this particular case, the S\,{\sc ii} line
profile does not follow that of Fe\,{\sc ii} at all, exhibiting an
inversely asymmetric shape which peaks at the velocity of the detected
Fe\,{\sc ii}$^\star$ absorption. Significant overabundance of
$\alpha$-elements and/or ionization effects might be present in this
system and bias the observed [S/Fe] ratio high. More generally, the
[S/H] and [S/Fe] ratios given in Table~\ref{tab:h2} should be
considered cautiously as they are strictly speaking upper limits to
the true metallicities and dust depletion factors. This makes the
result that both metallicities and dust depletion factors in the
sample of GRB-DLAs observed with UVES (and currently all other
high-resolution spectrographs) are usually modest, even stronger.
Interestingly, this is most clearly seen in the three systems
exhibiting the highest neutral hydrogen column densities, $\log
N($H$^0)\sim 22$, where [X/H$]<-1.5$ and [X/Fe$]<+0.2$. The nature of
these systems is discussed in Sect.~\ref{sec:conditions}.

The observed H$^0$ column density distribution of {\it Swift}-era
GRB-DLAs is relatively flat and not highly skewed towards extremely
high values \citep[see][and also
Table~\ref{tab:h2}]{2006A&A...460L..13J}. In addition, because there
is mounting evidence that the bulk of the H$^0$ gas in GRB-DLAs has a
galactic origin, with distances from the burst larger than 100~pc up
to kiloparsecs (see Sects.~\ref{sec:modeling} and \ref{sec:distance}),
the comparison of GRB-DLA properties to results drawn from QSO-DLA
samples must be relevant. This is corroborated by our finding that the
DLA cloud in the case of \grb\ is smooth and diffuse, and not
apparently perturbed by star-forming regions, with a broadening
parametre and physical size that are typical of the Galactic ISM. In
Fig.~\ref{fig:HImetdep}, we compare the H$^0$ column densities with
metallicities and dust depletion factors for the UVES GRB-DLA sample
with QSO-DLAs. Studies of H$_2$ in large QSO-DLA samples observed with
VLT/UVES \citep{2003MNRAS.346..209L,2008A&A...481..327N} have shown
that the presence of H$_2$ does not strongly depend on the total H$^0$
column density, and that the probability of finding H$_2$ with high
molecular fractions, $\log f>-4.5$, is only marginally higher for
$\log N($H$^0)>20.8$ than below this limit (19\% and 7\%
respectively). The probability of H$_2$ detection is of the order of
10\% for the entire QSO-DLA population, but is actually strongly
dependent on the metallicity, rising up to 35\% for systems with
[X/H$]\ge -1.3$ \citep{2006A&A...456L...9P,2008A&A...481..327N}. In
contrast, only about 4\% of the [X/H$]<-1.3$ systems have $\log
f>-4.5$. Applying this number statistics to the UVES GRB-DLA sample
(see Table~\ref{tab:h2}), composed of one [X/H$]>-1.3$ and four
[X/H$]<-1.3$ DLAs, leads to a binomial probability of detecting H$_2$
at least once of 45\%. This is consistent with the fact that no
detection was achieved in practice.

To estimate the dust content of DLAs, it is useful to calculate the
column density of iron in dust, $N($Fe$)_{\rm dust}=(1-10^{-[{\rm
    X}/{\rm Fe}]})N($X$)($Fe$/$X$)_{\rm dla}$
\citep{2006A&A...454..151V}. In their survey of 77 QSO-DLAs,
\citet{2008A&A...481..327N} have shown that all H$_2$-bearing systems
have $\log N($Fe$)_{\rm dust}>14.7$. In addition, about 17\% of the
systems with $14.7<\log N($Fe$)_{\rm dust}<15.2$, and 55\% of the
systems above this limit, have detected H$_2$ lines with $\log
f>-4.5$. The $\log N($Fe$)_{\rm dust}$ values pertaining to GRB-DLAs
are given in Table~\ref{tab:h2}. Except in the case of the
high-metallicity DLA towards GRB\,050820, which exhibits $\log
N($Fe$)_{\rm dust}\sim 16$, GRB-DLA values lie either within (in three
cases) or below (in one case) the above range. Therefore, from QSO-DLA
statistics the binomial probability of not detecting H$_2$ at all in
the UVES GRB-DLA sample is 26\%. From a $1.1\sigma$ result, relying on
the existence of a single system, it would be premature to conclude
that GRB-DLA observations are inconsistent with known QSO-DLA
properties. This is in contrast with the claim by
\citet{2007ApJ...668..667T}, based on an even smaller and
inhomogeneous sample of three high- and two medium-resolution GRB
afterglow spectra, that H$_2$ is generally deficient in GRB-DLAs
compared to QSO lines of sight.

\subsection{Gas physical conditions}
\label{sec:conditions}

The current lack of H$_2$ detection along the lines of sight to GRBs
might be related to the physical nature of DLAs exhibiting the highest
neutral hydrogen column densities, $\log N($H$^0)\sim 22$. Indeed,
these systems make up about half of the UVES GRB-DLA sample (i.e., the
systems towards \grb, GRB\,071031 and GRB\,080413A). Interestingly,
three systems in the large QSO-DLA sample studied by
\citet{2008A&A...481..327N} share similar properties with no detected
H$_2$ but some of the highest column densities of iron in dust ($\log
N($Fe$)_{\rm dust}>15.5$) as well as the highest neutral hydrogen
column densities ever measured along QSO lines of sight (i.e., $\log
N($H$^0)=21.70$ towards Q\,0458$-$0203, $\log N($H$^0)=21.80$ towards
Q\,1157$+$0128, and $\log N($H$^0)=21.40$ towards Q\,1209$+$0919). The
much smaller incidence rate of such DLAs in QSO samples is probably
due to the narrow range of galactocentric radii probed by GRB
afterglow observations. The properties of these systems are at
variance with Galactic lines of sight, where clouds with $\log
N($H$^0)>21$ usually have $\log N($H$_2)>19$ \citep[see,
e.g.,][]{1977ApJ...216..291S,1979ApJ...231...55J}.

\citet{2007ApJ...668..667T} argue that UV radiation fields from
recently formed hot stars with intensities 10-100 times the Galactic
mean value must be invoked to explain GRB-DLA observations. This is
conceivable if the clouds are embedded in or are located close to
regions of intense star formation \citep[see
also][]{2007ApJ...668..384C}. However, as noted by
\citet{2007ApJ...668..667T} this contrasts with the environment of
massively star-forming regions like 30~Doradus in the LMC where the
existence of clouds with high H$_2$ molecular fractions is ubiquitous
\citep{2001A&A...379...82B,2002ApJS..139...81D}. In other words, the
presence of intense UV radiation fields does not necessarily prevent
H$_2$ from forming.

In addition to a low dust content (see Sect.~\ref{sec:statistics}),
and the correspondingly low formation rate of H$_2$ onto dust grains,
another reason why H$_2$ is not detected in those QSO- and GRB-DLAs
with high H$^0$ column densities could be the particle density in the
neutral gas being too small. For a cloud at equilibrium, by equating
the H$_2$ formation ($R_{\rm dust}$) and photo-dissociation ($R_{\rm
  diss}$) rates, one can write \citep[see,
e.g.,][]{2005MNRAS.356.1529H}:

\begin{equation}
R_{\rm diss} N({\rm H_2}) = n_{\rm H} R_{\rm dust} N({\rm H})
\label{eq:h2eq}
\end{equation}

\noindent where $N($H$)$ is the total hydrogen column density and
$N($H$_2)$ the molecular hydrogen column density, which then linearly
depends on the particle density, $n_{\rm H}$. Therefore, the lower the
density the lower the molecular fraction. In the case of the DLA
towards \grb, from its high H$^0$ column density, $\log
N($H$^0)=22.10$, and considering
the best-fit linear cloud size of
$l=520^{+240}_{-190}$~pc, a meaningful order of magnitude estimate
of the particle density can be inferred: $n_{\rm H}\approx
5-15$~cm$^{-3}$. Using Eq.~\ref{eq:h2eq} and the prescriptions by
\citet{2005MNRAS.356.1529H}, we can estimate the UV intensity at $h\nu
= 12.87$~eV outside the \grb-DLA cloud, averaged over the solid angle,
taking into account H$_2$ self-shielding. In order to keep the amount
of H$_2$ below the observational threshold, $\log N($H$_2)<13.8$ (see
Table~\ref{tab:h2}), the UV intensity must be at least of the order of
the Galactic radiation field: $J_{\rm LW}\ga 3\times 10^{-20}$ erg
s$^{-1}$ cm$^{-2}$ Hz$^{-1}$ sr$^{-1}$. This differs by more than a
factor of ten from the results by Tumlinson et al. (see above). This
is because in the latter study a scaling of the Galactic H$_2$
formation rate based on the metallicity rather than on the dust-to-gas
ratio, $\kappa =10^{[{\rm X}/{\rm H}]}(1-10^{[{\rm Fe}/{\rm X}]})$,
was assumed or, equivalently, all metals are incorporated into dust
grains. However, a strong correlation between depletion factor and
metallicity is observed in QSO-DLAs \citep[see,
e.g.,][]{2003MNRAS.346..209L}, with smaller depletion factors in low
metallicity systems, indicating that the dust-to-metal ratio increases
markedly in the course of chemical evolution
\citep{2004A&A...421..479V}.

While, as shown above, particle densities are probably at most
moderate in GRB-DLAs, with $n_{\rm H}\approx 5-15$~cm$^{-3}$, the
ambient UV radiation field intensities needed to explain the lack of
H$_2$ in systems with high H$^0$ column densities are only of the
order of 1-10 times those observed in the Galaxy, thus not
particularly high. As shown in Sect.~\ref{sec:statistics}, the main
factor driving the absence of H$_2$ in the current sample of GRB-DLAs
observed with VLT/UVES is a combination of low metallicities and low
dust contents.

\section{Summary and prospects}
\label{sec:conclusions}

In this paper, we have presented a detailed analysis of the DLA cloud
($\log N($H$^0)=22.10$) at the \grb\ host-galaxy redshift of $z_{\rm
  abs}=3.969$ using high-quality VLT/UVES spectra. Accurate metal
abundances are derived showing that this GRB-DLA has both low
metallicity, [S/H$]=-2.18\pm 0.11$, and no dust. In addition, we
measured [N/S$]=-0.98\pm 0.05$ and [Ar/S$]=-0.11\pm 0.06$. The former
ratio is indicative of a high nitrogen enrichment dominated by
intermediate-mass stars ($4-8$~$M_\odot$), before significant oxygen
production by supernovae, and a past low star-formation efficiency
\citep[e.g.,][]{2007PASP..119..962H,2008A&A...480..349P}. The latter
ratio is a tracer of the radiation field in H$^0$ regions
\citep{2003A&A...402..487V}. The observed high abundance of argon,
typical of the local ISM, is consistent with the expectation for a
neutral medium embedded in a soft ionizing continuum.

From the self-consistent photo-excitation modelling of absorption
lines from electronic transitions to an unprecedentedly large number
of Fe$^+$ energy levels, the distance of the absorbing neutral
material to \grb\ is constrained to be $d\approx 0.5$~kpc. Similar
photo-excitation modelling applied to the observed column densities
and upper limits of S$^{3+}$ energy levels has led to a lower limit on
the distance of the S$^{3+}$-bearing gas of 0.4~kpc
\citep{2008A&A...491..189F}. From detailed calculations, we find that
at such distances GRB afterglow photons cannot ionize H$^0$ nor
dissociate H$_2$ significantly. From the above Fe$^+$ column density
modelling, a linear cloud size of $l=520^{+240}_{-190}$~pc is
inferred, suggesting that GRB-DLAs typically exhibiting very high
H$^0$ column densities are diffuse metal-poor atomic clouds with large
physical extents, $l\ga 100$~pc, and, at most, moderate particle
densities, $n_{\rm H}\approx 5-15$~cm$^{-3}$.

In order to understand the lack of H$_2$ further, we have built up the
most up-to-date sample of {\it Swift}-era $z>1.8$ GRB-DLAs observed
with VLT/UVES. We showed that the current non-detections of H$_2$ in
GRB-DLAs are consistent with QSO-DLA statistics where H$_2$ is present
with high molecular fractions, $\log f>-4.5$, in only 10\% of the
global QSO-DLA population. Moreover, the lack of H$_2$ in the GRB-DLA
sample can be entirely explained by the low metallicities, [X/H$]<-1$,
and low dust contents of the systems. With the possible exception of
the high-metallicity DLA towards GRB\,050820, there is no need for
enhanced UV radiation fields from recently formed hot stars to explain
the lack of H$_2$ (but see \citealt{2007ApJ...668..667T}; see also
\citealt{2009ApJ...691..152C} who derived high mean UV radiation field
intensities for four bright GRB host galaxies, whose associated DLAs
all have fairly high metallicities). The additional non-detection of
C\,{\sc i} absorption lines, with $\log N($C$^0)/N($S$^+)<-3$,
suggests a warm neutral medium, with kinetic temperatures $T_{\rm
  kin}\ga 1000$~K, as found in most QSO-DLAs
\citep{2005MNRAS.362..549S}.

The evidence for low to very low GRB-DLA metallicities coming from
high spectral resolution observations is in contrast with the current
sample of metallicity and depletion measurements drawn from low-
and/or intermediate-resolution spectroscopy
\citep[e.g.,][]{2006A&A...451L..47F,2006NJPh....8..195S,2007ApJ...666..267P}
which points towards higher metallicities than in the QSO-DLA
population \citep[see also][who tried to reconcile the different
metallicity distributions of QSO- and GRB-DLAs]{2008ApJ...683..321F}.
Moreover, hints at the presence of H$_2$ and/or C\,{\sc i} lines have
been reported at the host-galaxy redshifts of GRB\,050401
\citep{2006ApJ...652.1011W}, GRB\,060206 \citep{2006A&A...451L..47F},
and GRB\,070802 \citep{2009ApJ...697.1725E}, while the metallicities
of these hosts are amongst the highest measured in GRB-DLAs. This
suggests that it is only a matter of time before H$_2$ and/or other
molecules are detected beyond any doubt in GRB-DLAs\footnote{Since the
  submission of this paper, CO molecules with exceptionally high
  column densities have been detected at the host-galaxy redshift of
  GRB\,080607 in a low-resolution Keck afterglow spectrum
  \citep{2009ApJ...691L..27P}. The absorbing gas is estimated to have
  roughly solar metallicity.}. Care should be exercized, however, not
to overinterpret the current inhomogeneous body of data, as claims for
high Zn$^+$ column densities and depletion factors of zinc compared to
iron are based on a handful of absorbers at intermediate redshifts,
$1<z<2$, for which $N($H$^0)$ measurements are lacking
\citep{2006NJPh....8..195S}.

The VLT/UVES sample of high-redshift GRB afterglows shows a lack of
GRB-DLAs with both high H$^0$ column density and high metallicity (see
Fig.~\ref{fig:HImetdep}, left panel). However, this does not mean that
such systems do not exist and an anti-correlation between absorber
dust content and afterglow optical brightness should be checked.
Indeed, the associated extinction will be high in these systems
(\citealt{2005A&A...444..461V}; see also
\citealt{2009ApJ...691L..27P}) and their observation with
high-resolution spectrographs, even with the advantage of fast
response, possibly difficult. This is consistent with the findings of
multi-colour photometry that a large fraction of bursts are dark
because of high dust extinction in their hosts
\citep{2007ApJ...669.1098R,2008ApJ...681..453J,2008MNRAS.388.1743T,2009ApJ...693.1484C}.
If a bias exists in the UVES sample, it could also be against GRB-DLAs
located at much smaller distances from their bursts than currently
found because, as we have shown, high metal (resp. H$_2$) column
densities are able to survive the ionizing (resp. dissociating)
effects of the incident GRB afterglow radiation.

This potential problem should be alleviated by the second-generation
VLT instrument X-Shooter, currently in its commissioning phase, to
complement the capabilities of UVES at the VLT, Unit~2 -- Kueyen,
telescope. X-Shooter will be the instrument of choice to cover any
redshift from $z=0.3$ up to $z=10$ for 1600~\AA\ (rest frame) Fe\,{\sc
  ii} and other metal lines in GRB afterglows fainter, or redder, than
observable with UVES. In the meantime, we plan to apply the techniques
presented in this paper for \grb\ to other GRB lines of sight observed
with UVES to start building up a sample of burst/DLA distance
determinations and map the distribution of neutral gas in GRB host
galaxies.

\begin{acknowledgements}
  PMV acknowledges the support of the EU under a Marie Curie
  Intra-European Fellowship, contract MEIF-CT-2006-041363. The Dark
  Cosmology Centre is funded by the Danish National Research
  Foundation.
\end{acknowledgements}

\bibliographystyle{aa}
\bibliography{11572}

\end{document}